\documentclass[letterpaper,english,reprint, aps]{revtex4-1}
\usepackage[T1]{fontenc}
\usepackage[latin9]{luainputenc}
\usepackage{amsmath}
\usepackage{amssymb}
\usepackage{graphicx}

\makeatletter

\@ifundefined{pageheight}{\let\pageheight\pdfpageheight}{}
\@ifundefined{pagewidth}{\let\pagewidth\pdfpagewidth}{}
\pageheight\paperheight
\pagewidth\paperwidth

\makeatother

\usepackage{babel}
\begin{document}

\preprint{APS/123-QED}

\title{Light-cone like spreading of single-particle correlations in the
Bose-Hubbard model after a quantum quench in the strong coupling regime}

\author{Matthew R.~C. Fitzpatrick}
\email{mrfitzpa@sfu.ca}

\selectlanguage{english}%

\affiliation{Department of Physics, Simon Fraser University,~\\
 8888 University Drive, Burnaby, British Columbia V5A 1S6, Canada
}

\author{Malcolm P. Kennett}
\email{malcolmk@sfu.ca}

\selectlanguage{english}%

\affiliation{Department of Physics, Simon Fraser University,~\\
 8888 University Drive, Burnaby, British Columbia V5A 1S6, Canada
}

\date{\today}
\begin{abstract}
We study the spreading of correlations in space and time after a quantum quench in the Bose
Hubbard model. We derive equations of motion for the single-particle
Green's function within the contour-time formalism, allowing us
to study dynamics in the strong coupling regime. We discuss the numerical
solutions of these equations and calculate the single-particle density
matrix for quenches in the Mott phase. We demonstrate light-cone like
spreading of correlations in the Mott phase in one, two, and three
dimensions and calculate propagation velocities in each dimension. 
\end{abstract}



\maketitle

\section{\label{sec:introduction}Introduction}

The out-of-equilibrium dynamics of interacting quantum systems has
become a major subject of interest in many-body physics. Experimental
advances have made ultracold atoms in optical lattices offer a promising 
setting to study  out-of-equilibrium phenomena and attracted considerable 
attention in recent years \citep{BlochRev1,BlochRev2,Morsch,Lewenstein,Bloch,KennettRev}.
These systems are highly versatile in that experimental parameters
can be tuned over a wide range of values in real time. This 
facilitates the study of quantum quenches, in which parameters in
the corresponding Hamiltonian are varied in time faster than the system
can respond adiabatically. Such protocols open the door to a rich
range of many-body physics and have been studied intensely both theoretically
and experimentally. 

Jaksch \emph{et al. }\citep{Jaksch} showed that ultracold bosons trapped 
in optical lattices can be described by the Bose-Hubbard model (BHM) -- a 
minimal model of interacting bosons on a lattice.
The BHM exhibits a quantum phase transition between a superfluid and
Mott-insulator as the ratio of the hopping strength, $J$, to the on-site
interaction strength, $U$, is varied \citep{Fisher}, which was demonstrated experimentally
for cold atoms by Greiner \emph{et al. }\citep{Greiner}. This allows for 
the study of quantum quenches across a quantum critical point, in addition 
to quenches within a particular phase. 

A variety of quench protocols have been suggested 
and implemented \citep{Greiner,Chen,Hung,Bakr} for the BHM in order to study out of equilibrium 
phenomena such as the Kibble-Zurek effect \cite{Kibble,Zurek,Zurek05,Chen} and relaxation
after a quench \citep{Clark,Kollath,Sciolla,Sciolla2,Fischer0,Fischer1,Kennett,Werner,Nessi,Polkovnikov_nex,Natu,Bernier,Natu2,Bernier2,Zakrzewski,Trefzger,Yanay}.
Our particular interest here is the light-cone like spreading of correlations
after a quantum quench. Several analytical and numerical studies have
shown a Lieb-Robinson-like \cite{LiebRobinson} maximal propagation velocity for the spreading of density correlations
in one dimensional systems for quenches from the superfluid to Mott-insulating
regime as well as quenches solely within the superfluid \citep{Carleo}
or Mott-insulating regimes \citep{Fischer0,Bernier,Lauchli,Barmettler,Krutitsky}.
The latter case was recently observed by Cheneau \emph{et al. }\citep{Cheneau}
for an array of decoupled one-dimensional chains.  Some theoretical predictions have
also been made for higher dimensional systems \citep{Carleo,Navez,Natu2,Krutitsky}
but these have not yet faced experimental scrutiny.

A generic problem in the theoretical description of quantum quenches is that
it is necessary to have a formalism that is able to describe the physics in a broad
area of parameter space.  In the case of the Bose Hubbard model,
numerical approaches such as exact diagonalization (ED) and
the time-dependent density matrix renormalization
group (t-DMRG) \cite{Kollath,Lauchli,Bernier,Bernier2,Clark,Cheneau,Trotzky1d}
can be essentially exact in all parts of parameter space but are limited by system size
and usually are practical only in one dimension. For dimensions higher than one, 
methods such as time-dependent Gutzwiller mean field theory \cite{Lewenstein,Zakrzewski,Amico,Natu} 
and dynamical mean field theory \cite{Werner} have been used which can capture the presence 
of a quantum phase transition, but in their simplest form do not capture spatial
correlations.  However, there has been work on including 
perturbative corrections \cite{Trefzger,TrefzgerDutta,Navez,Schroll,Yanay,Navez2,Krutitsky} 
to remedy this weakness. 

In previous work \citep{Fitzpatrick}, we developed a real-time two-particle
irreducible (2PI) effective action approach to the BHM based on a 
 strong-coupling theory of the BHM \citep{SenguptaDupuis,Kennett} that is exact in 
 both the weak and strong coupling limits. 
 We verified that by using a Hartree-Fock-Bogoliubov approximation we were 
able to obtain considerable improvements beyond mean field theory
in calculating equilibrium properties of the BHM \citep{Fitzpatrick}.  We also
derived  equations of motion for the single-particle Green's function 
using the contour-time formalism \citep{KonstantinovPerel}.  In this paper
we use the equations of motion to investigate the case of a quench in the 
Mott-insulating regime. We demonstrate light-cone spreading of 
single-particle correlations in one, two and three dimensions.  
We also study the dependence of the maximal propagation velocity
on quench protocol, chemical potential, temperature and dimensionality
that should be 
relevant for comparisons with experiment.

The paper is structured as follows. In Sec.~\ref{sec:Model and formalism},
we describe the model that we study and the theoretical formalism
we use to calculate correlations after a quench. In Sec.~\ref{sec:Equations of motion},
we briefly discuss the equations of motion for the single-particle
Green's function that we obtained in our previous work \citep{Fitzpatrick}
and show how they simplify for quenches that are confined to the Mott
regime. In Sec.~\ref{sec:Numerical results}, we present numerical
results obtained from integrating the quations of motion and finally
in Sec.~\ref{sec:Discussion and conclusions}, we discuss our results
and present our conclusions.

\section{\label{sec:Model and formalism}Model and Formalism}

In this section we introduce the Bose-Hubbard model and the
effective theory we use to study quench dynamics in the strong-coupling
regime, all within the context of the contour-time formalism. The
Hamiltonian for the BHM, allowing for a time dependent hopping term,
is

\begin{equation}
\hat{H}_{\text{BHM}}\left(t\right)=\hat{H}_{J}\left(t\right)+\hat{H}_{0},\label{eq:BHM Hamiltonian - 1}
\end{equation}

\noindent where

\begin{equation}
\hat{H}_{J}\left(t\right)=-\sum_{\left\langle \vec{r}_{1},\vec{r}_{2}\right\rangle }J_{\vec{r}_{1}\vec{r}_{2}}\left(t\right)\left(\hat{a}_{\vec{r}_{1}}^{\dagger}\hat{a}_{\vec{r}_{2}}+\hat{a}_{\vec{r}_{2}}^{\dagger}\hat{a}_{\vec{r}_{1}}\right),\label{eq:H_J defined - 1}
\end{equation}
and
\begin{equation}
\hat{H}_{0}=\hat{H}_{U}-\mu\hat{N}=\frac{U}{2}\sum_{\vec{r}}\hat{n}_{\vec{r}}\left(\hat{n}_{\vec{r}}-1\right)-\mu\sum_{\vec{r}}\hat{n}_{\vec{r}},\label{eq:H_0 defined - 1}
\end{equation}

\noindent with $\hat{a}_{\vec{r}}^{\dagger}$ and $\hat{a}_{\vec{r}}$
annihilation and creation operators for bosons on lattice site $\vec{r}$
respectively, $\hat{n}_{\vec{r}}\equiv\hat{a}_{\vec{r}}^{\dagger}\hat{a}_{\vec{r}}$
the number operator, $U$ the interaction strength, and $\mu$ the
chemical potential. The notation $\left\langle \vec{r}_{1},\vec{r}_{2}\right\rangle $
indicates a sum over nearest neighbours only. We allow $J_{\vec{r}_{1}\vec{r}_{2}}\left(t\right)$,
the hopping amplitude between sites $\vec{r}_{1}$ and $\vec{r}_{2}$,
to be time dependent.  We have specified the model for a uniform lattice,
but could consider a trap as is used in experiment by introducing a site-dependent chemical potential.
This leads to more complicated calculations than we consider here
but is conceptually straightforward to include.

\subsection{\label{subsec:Contour-time formalism}Contour-time formalism}

\begin{figure}[h]
\begin{centering}
\includegraphics[scale=0.6]{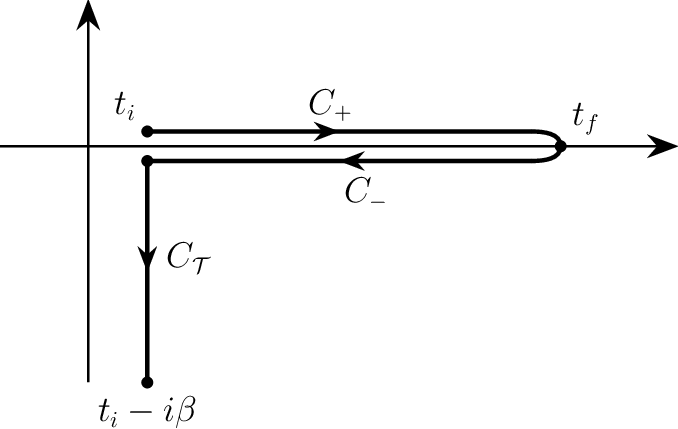}
\par\end{centering}
\caption{Contour for a system initially prepared at time $t_{i}$ in a thermal
state with inverse temperature $\beta$. $t_{f}$ is the maximum real-time
considered in the problem, which may be set to $t_{f}\to\infty$ without
loss of generality.\label{fig:fig1}}
\end{figure}

The general formalism that we discuss and adopt in this paper was
developed in a previous paper of ours; we refer the reader to Ref.~\citep{Fitzpatrick}
for details on the formalism. We use the contour-time formalism \citep{Schwinger,Keldysh,Rammer,Semenoff,Landsman,Chou},
which treats time as a complex variable lying along a contour in 
a way that allows the description of out-of-equilibrium and equilibrium
quantum phenomena within the same formalism. For systems initially
prepared in thermal states, which we consider here, one can work with
a contour $C$ of the form illustrated in Fig.~\ref{fig:fig1} which
is sometimes referred to as the Konstantinov and Perel' (KP) contour
\citep{KonstantinovPerel}. A popular alternative to the KP contour
is the Schwinger-Keldysh (SK) closed-time path \citep{Schwinger,Keldysh}
which is also suitable for initially thermalized systems. However,
unlike the KP contour, the SK contour ignores transient phenomena,
being more suitable for calculating steady states or other long-time
phenomena. Given that transient effects are important in the spreading
of space-time correlations after a quantum quench, the KP contour
is a more appropriate choice. A number of authors have applied contour-time
approaches to the BHM in out-of-equilibrium scenarios \citep{Robertson,Dalidovich,Fitzpatrick,Kennett,Gras,Grass2,Grassthesis,Rey1,Rey2,Temme,Calzetta,PolkovnikovCTP,DellAnna}
\textendash{} our work differs from previous approaches \citep{Rey1,Temme} in that we
apply an effective theory to the BHM within the contour formalism
that is appropriate for strong coupling as well as weak coupling \citep{Kennett,Fitzpatrick}. 

\subsection{\label{subsec:Contour-ordered Green's functions}Contour-ordered
Green's functions}

To characterize spatio-temporal correlations in the BHM we calculate
contour-ordered Green's functions (COGFs). We define the $n$-point
COGF as \citep{Chou}
\begin{widetext}
\begin{align}
G_{\vec{r}_{1}\ldots\vec{r}_{n}}^{a_{1}\ldots a_{n}}\left(\tau_{1},\ldots,\tau_{n}\right) & \equiv\left(-i\right)^{n-1}\text{Tr}\left\{ \hat{\rho}_{i}T_{C}\left[\hat{a}_{\vec{r}_{1}}^{a_{1}}\left(\tau_{1}\right)\ldots\hat{a}_{\vec{r}_{n}}^{a_{n}}\left(\tau_{n}\right)\right]\right\} \nonumber \\
 & \equiv\left(-i\right)^{n-1}\left\langle T_{C}\left[\hat{a}_{\vec{r}_{1}}^{a_{1}}\left(\tau_{1}\right)\ldots\hat{a}_{\vec{r}_{n}}^{a_{n}}\left(\tau_{n}\right)\right]\right\rangle _{\hat{\rho}_{i}},\label{eq:COGFs defined - 1}
\end{align}
\end{widetext}

\noindent where $\hat{\rho}_{i}$ is the state operator for a thermal
state representing the initial state of our system:

\begin{equation}
\hat{\rho}_{i}=\frac{e^{-\beta\hat{H}_{\text{BHM}}\left(t_{i}\right)}}{\text{Tr}\left\{ e^{-\beta\hat{H}_{\text{BHM}}\left(t_{i}\right)}\right\} },\label{eq:State operator for initial thermal state - 1}
\end{equation}

\noindent the $a_{i}$ upper indices are defined such that

\begin{equation}
\hat{a}_{\vec{r}}^{1}\equiv\hat{a}_{\vec{r}},\quad\hat{a}_{\vec{r}}^{2}\equiv\hat{a}_{\vec{r}}^{\dagger},\label{eq:introducing nambu indices - 1}
\end{equation}

\noindent and $\hat{a}_{\vec{r}}^{a}\left(\tau\right)$ are the bosonic
fields in the Heisenberg picture with respect to $\hat{H}_{\text{BHM}}\left(\tau\right)$
{[}Eq.~\eqref{eq:BHM Hamiltonian - 1}{]}:

\begin{align}
\hat{a}_{\vec{r}}^{a}\left(\tau\right) & =U^{\dagger}\left(\tau,\tau_{i}\right)\hat{a}_{\vec{r}}^{a}U\left(\tau,\tau_{i}\right),\label{eq:Heisenberg fields - 1}\\
U\left(\tau,\tau^{\prime}\right) & =T_{C}\left[e^{-\int_{C\left(\tau,\tau^{\prime}\right)}d\tau^{\prime\prime}\hat{H}_{\text{BHM}}\left(\tau^{\prime\prime}\right)}\right].\label{eq:evolution operator - 1}
\end{align}

\noindent Here we have introduced explicitly the complex contour time
argument $\tau$, the sub-contour $C\left(\tau,\tau^{\prime}\right)$
which goes from $\tau$ to $\tau^{\prime}$ along the contour $C$,
and the contour time ordering operator $T_{C}$, which orders strings
of operators according to their position on the contour, with operators
at earlier contour times placed to the right.

\subsection{\label{subsec:Effective theory for the BHM}Effective theory for
the Bose-Hubbard model}

In order to study quench dynamics in the BHM, we make use of an effective
theory (expressed as an action) that can describe both the weak and
strong coupling limits of the model in the same formalism. Such an
approach was developed in imaginary time by Sengupta and Dupuis \citep{SenguptaDupuis}
by using two Hubbard-Stratonovich transformations, then generalized
to the SK contour in Ref.~\citep{Kennett}, and then further generalized
to the KP contour in Ref.~\citep{Fitzpatrick} in conjunction with
a 2PI effective action approach. A similar real-time theory was also
obtained based on a Ginzburg-Landau approach using the Schwinger-Keldysh
technique \citep{Gras,Grass2,Grassthesis} in conjunction with a one-particle
irreducible (1PI) effective action approach. In obtaining the effective
theory below, one assumes that the system is dominated by the low-energy
degrees of freedom, which is valid as along as the quench is sufficiently
slow. A detailed discussion of the development of the effective theory
within the KP contour formalism is presented in Ref.~\citep{Fitzpatrick}.
The effective theory obtained in Ref.~\citep{Fitzpatrick} for
$z$ fields (which are obtained after two Hubbard Stratonovich transformations
and have the same correlations as the original $a$ fields \citep{SenguptaDupuis})
is
\begin{widetext}
\begin{align}
S\left[z\right] & =\frac{1}{2!}\sum_{\vec{r}}\int_{C}\int_{C}d\tau_{1}d\tau_{2}\left[\mathcal{G}^{-1}\right]^{a_{1}a_{2}}\left(\tau_{1},\tau_{2}\right)z_{\vec{r}}^{\overline{a_{1}}}\left(\tau_{1}\right)z_{\vec{r}}^{\overline{a_{2}}}\left(\tau_{2}\right)\nonumber \\
 & \quad+\frac{1}{2!}\sum_{\vec{r}_{1}\vec{r}_{2}}\int_{C}d\tau\left\{ 2J_{\vec{r}_{1}\vec{r}_{2}}\left(\tau\right)+\delta_{\vec{r}_{1}\vec{r}_{2}}v_{1}\right\} \sigma_{1}^{a_{1}a_{2}}z_{\vec{r}_{1}}^{\overline{a_{1}}}\left(\tau\right)z_{\vec{r}_{2}}^{\overline{a_{2}}}\left(\tau\right)\nonumber \\
 & \quad+\frac{1}{4!}\sum_{\vec{r}}\int_{C}d\tau\left\{ -2u_{1}\right\} \sigma^{a_{1}a_{2}a_{3}a_{4}}z_{\vec{r}}^{\overline{a_{1}}}\left(\tau\right)z_{\vec{r}}^{\overline{a_{2}}}\left(\tau\right)z_{\vec{r}}^{\overline{a_{3}}}\left(\tau\right)z_{\vec{r}}^{\overline{a_{4}}}\left(\tau\right),\label{eq:effective theory - 1}
\end{align}
\end{widetext}

\noindent where $\mathcal{G}^{a_{1}a_{2}}\left(\tau_{1},\tau_{2}\right)$
is the atomic (i.e. $J=0$) two-point Green's function (see Appendix
C of Ref.~\citep{Fitzpatrick} for the full expression), $u_{1}$ is a complicated
function of the inverse temperature $\beta$ and the chemical potential
$\mu$ (see Appendix D of Ref.~\citep{Fitzpatrick} for the full expression), and

\begin{align}
v_{1} & =\left(2n_{J=0}+1\right)u_{1},\label{eq:v1 - 1}
\end{align}

\noindent where $n_{J=0}$ is the average particle density in the atomic
limit.  This can be calculated from the atomic kinetic Green's function
$\mathcal{G}^{12,\left(K\right)}$ (see Appendix C of Ref.~\citep{Fitzpatrick})
as follows

\begin{align}
n_{J=0} & =\frac{1}{2}\left\{ i\mathcal{G}_{\vec{k}}^{12,\left(K\right)}\left(t^{\prime}=0\right)-1\right\} .\label{eq:atomic particle density - 1}  
\end{align}

\noindent the overscored index $\overline{a}$ used in Eq.~\eqref{eq:effective theory - 1} is defined by

\begin{equation}
f_{\vec{r}}^{\overline{a}}\left(\tau\right)\equiv\sigma_{1}^{aa^{\prime}}f_{\vec{r}}^{a^{\prime}}\left(\tau\right),\label{eq:nambu summation convention - 1}
\end{equation}

\noindent where $\sigma_{i}$ is the $i^{\text{th}}$ Pauli matrix,
i.e. $\overline{1}=2$ and $\overline{2}=1$, and 

\begin{equation}
\sigma^{a_{1}a_{2}a_{3}a_{4}}\equiv\begin{cases}
1, & \text{if }\left\{ a_{m}\right\} _{m=1}^{4}\in P\left(\left\{ 1,1,2,2\right\} \right)\\
0, & \text{otherwise}
\end{cases}.\label{eq:sigma 4 - 1}
\end{equation}

\noindent We use the Einstein summation convention for the Nambu indices,
i.e. matching indices implies a summation over all possible values
of those indices. 

When applied to an nPI effective action approach, where one 
ultimately calculates equations of motion for various different correlation
functions, the effective theory generates 
``anomalous'' Feynman diagrams \citep{Dupuis,SenguptaDupuis,Kennett,Fitzpatrick,Sajna}.
These diagrams contain internal inverse atomic propagator lines which
do not correspond to any physical processes.  If one considers all orders of 
the theory, they can be dropped because the different anomalous terms cancel.
If the theory is truncated (as is usually the case), then care is required to
ensure cancellation order by order.  At the level considered here, the $v_1$ 
term in  Eq.~\eqref{eq:effective theory - 1} plays this role. For
a more detailed discussion of the cancellation of anomalous diagrams,
see Ref.~\citep{Fitzpatrick}.

The effective theory introduces an effective potential $v_{1}$ and
a renormalized on-site interaction strength $u_{1}$. Moreover, it
reassigns the role of the ``bare propagator'' to the atomic propagator.
The theory gives the exact
two-point connected COGF (CCOGF) in both the atomic and noninteracting
limits, thus making it particularly appealing for the study of quench
dynamics since it gives a reasonable description of
the behaviour of the system in both the superfluid and Mott-insulating
regimes \citep{KennettRev}.

\section{\label{sec:Equations of motion}Equations of motion}

Our goal is to calculate the full two-point CCOGF
(the ``full propagator'' from now on) after a quench, which encodes
non-local single-particle spatial and temporal correlations. To achieve
this, we solve the Dyson's equation \citep{Cornwall,Fitzpatrick} for the 
full propagator (the superscript ``c'' indicates that $G$ is a connected COGF):
\begin{widetext}
\begin{equation}
G_{\vec{k}}^{a_{1}a_{2},c}\left(\tau_{1},\tau_{2}\right)\equiv\left[G_{0}\right]_{\vec{k}}^{a_{1}a_{2},c}\left(\tau_{1},\tau_{2}\right)+\int_{C}\int_{C}d\tau_{3}d\tau_{4}\left[G_{0}\right]_{\vec{k}}^{a_{1}a_{3},c}\left(\tau_{1},\tau_{3}\right)\Sigma_{\vec{k}}^{\overline{a_{3}}\overline{a_{4}}}\left(\tau_{3},\tau_{4}\right)G_{\vec{k}}^{a_{4}a_{2},c}\left(\tau_{4},\tau_{2}\right),\label{eq:Dyson's equation - 1}
\end{equation}
\end{widetext}

\noindent where $G_{0}$ is the bare propagator and $\Sigma$ is the
self-energy of the theory. Since we consider
a translationally invariant system, we work in quasi-momentum space rather
than real space. In Ref.~\citep{Fitzpatrick},
we calculated the self-energy for the effective theory {[}Eq.~\eqref{eq:effective theory - 1}{]}
in a systematic way using a 2PI effective action approach \citep{Cornwall}
and considered terms up to first order in $u_{1}$ (loosely corresponding
to a Hartree-Fock-Bogoliubov (HFB) like approximation).

The equations of motion derived in Ref.~\citep{Fitzpatrick} are
quite general in that they can be applied to a variety of different
quench protocols. Here we consider the case in which the hopping quench is
restricted to the Mott-insulating regime and the system is initially
thermalized in the atomic limit. Under these conditions, the self-energy
(and thus the equations of motion) simplify considerably, and it is
straightforward to show that the equations of motion derived in Ref.~\citep{Fitzpatrick}
reduce to 
\begin{widetext}
\begin{align}
A_{\vec{k}}\left(t,t^{\prime}\right) & =\mathcal{A}\left(t-t^{\prime}\right) -i\int_{t^{\prime}}^{t}dt^{\prime\prime}\mathcal{A}\left(t-t^{\prime\prime}\right)\Sigma_{\vec{k}}^{\left(HFB\right)}\left(t^{\prime\prime}\right)A_{\vec{k}}\left(t^{\prime\prime},t^{\prime}\right),\label{eq:A eqn of motion - 1}\\
G_{\vec{k}}^{\left(K\right)}\left(t,t^{\prime}\right) & =\mathcal{G}^{\left(K\right)}\left(t-t^{\prime}\right) -i\int_{0}^{t}dt^{\prime\prime}\mathcal{A}\left(t-t^{\prime\prime}\right)\Sigma_{\vec{k}}^{\left(HFB\right)}\left(t^{\prime\prime}\right)G_{\vec{k}}^{\left(K\right)}\left(t^{\prime\prime},t^{\prime}\right)\nonumber \\
 & \quad+i\int_{0}^{t^{\prime}}dt^{\prime\prime}\mathcal{G}^{\left(K\right)}\left(t-t^{\prime\prime}\right)\Sigma_{\vec{k}}^{\left(HFB\right)}\left(t^{\prime\prime}\right)A_{\vec{k}}\left(t^{\prime\prime},t^{\prime}\right),\label{eq:G_K eqn of motion - 1}
\end{align}
\end{widetext}

\noindent where $A_{\vec{k}}\left(t,t^{\prime}\right)$ is the spectral
function:

\begin{equation}
A_{\vec{k}}\left(t,t^{\prime}\right)=\left\langle \hat{a}_{\vec{k}}\left(t\right)\hat{a}_{\vec{k}}^{\dagger}\left(t^{\prime}\right)-\hat{a}_{\vec{k}}^{\dagger}\left(t^{\prime}\right)\hat{a}_{\vec{k}}\left(t\right)\right\rangle _{\hat{\rho}_{i}},\label{eq:spectral func def - 1}
\end{equation}
\noindent and  $G_{\vec{k}}^{\left(K\right)}\left(t,t^{\prime}\right)$
is the kinetic Green's function:

\begin{align}
G_{\vec{k}}^{\left(K\right)}\left(t,t^{\prime}\right) & =G_{\vec{k}}^{12,\left(K\right)}\left(t,t^{\prime}\right)\nonumber \\
 & =-i\left\langle \hat{a}_{\vec{k}}\left(t\right)\hat{a}_{\vec{k}}^{\dagger}\left(t^{\prime}\right)+\hat{a}_{\vec{k}}^{\dagger}\left(t^{\prime}\right)\hat{a}_{\vec{k}}\left(t\right)\right\rangle _{\hat{\rho}_{i}}.\label{eq:kinetic green's func def - 1}
\end{align}

\noindent The quantities  $\mathcal{A}\left(t-t^{\prime}\right)$ and $\mathcal{G}^{\left(K\right)}\left(t-t^{\prime}\right)$
that enter Eqs.~\eqref{eq:A eqn of motion - 1} and \eqref{eq:G_K eqn of motion - 1} are the spectral function in the atomic limit
 and the kinetic Green's function in the atomic limit respectively.  In this limit both quantities are time-translational invariant.
$\Sigma_{\vec{k}}^{\left(HFB\right)}\left(t\right)$ is the self-energy in the HFB approximation:

\begin{equation}
\Sigma_{\vec{k}}^{\left(HFB\right)}\left(t\right)=\epsilon_{\vec{k}}\left(t\right)+2u_{1}\left\{ n\left(t\right)-n_{J=0}\right\} ,\label{eq:HFB self energy - 1}
\end{equation}

\noindent with 

\begin{align}
\epsilon_{\vec{k}}\left(t\right) & =-2J\left(t\right)\sum_{i=1}^{d}\cos\left(k_{i}a\right),\label{eq:epsilon - 1}\\
n\left(t\right) & =\frac{1}{N_{\text{sites}}}\sum_{\vec{k}}n_{\vec{k}}\left(t\right),\label{eq:particle density - 1}\\
n_{\vec{k}}\left(t\right) & =\frac{1}{2}\left\{ iG_{\vec{k}}^{\left(K\right)}\left(t,t\right)-1\right\} ,\label{eq:n_k - 1}
\end{align}

\noindent and $a$ the lattice constant (assuming a $d$-dimensional
hypercube geometry). In the atomic limit, the spectral function and
kinetic Green's functions can be written as
\begin{widetext}
\begin{align}
\mathcal{A}\left(t\right) & =\frac{1}{\mathcal{Z}}\sum_{n=0}^{\infty}e^{-\beta\left(\mathcal{E}_{n}-\mathcal{E}_{n_{\text{MI}}}\right)}\left\{ \left(n+1\right)e^{-i\left(\mathcal{E}_{n+1}-\mathcal{E}_{n}\right)t}-ne^{i\left(\mathcal{E}_{n-1}-\mathcal{E}_{n}\right)t}\right\} ,\label{eq:local spectral func - 1}\\
\mathcal{G}^{\left(K\right)}\left(t\right) & =-\frac{i}{\mathcal{Z}}\sum_{n=0}^{\infty}e^{-\beta\left(\mathcal{E}_{n}-\mathcal{E}_{n_{\text{MI}}}\right)}\left\{ \left(n+1\right)e^{-i\left(\mathcal{E}_{n+1}-\mathcal{E}_{n}\right)t}+ne^{i\left(\mathcal{E}_{n-1}-\mathcal{E}_{n}\right)t}\right\} ,\label{eq:local kinetic green's func - 1}
\end{align}
\end{widetext}

\noindent where $\mathcal{E}_{n}$ is the single-site energy:

\begin{equation}
\mathcal{E}_{n}=\frac{U}{2}n\left(n-1\right)-\mu n,\label{eq:single-site energy atomic limit - 1}
\end{equation}

\noindent $n_{\text{MI}}$ is the zero-temperature particle density:

\begin{equation}
n_{\text{MI}}=\left\lceil \mu/U\right\rceil ,\label{eq:n_MI - 1}
\end{equation}

\noindent and $\mathcal{Z}$ is the partition function:

\begin{equation}
\mathcal{Z}=\sum_{n=0}^{\infty}e^{-\beta\left(\mathcal{E}_{n}-\mathcal{E}_{n_{\text{MI}}}\right)}.\label{eq:local partition func - 1}
\end{equation}

We consider quenches in which the hopping amplitude $J\left(t\right)$
is tuned as a function of time. {[}Experimentally this corresponds
to varying the depth of the optical lattice,
since hopping varies exponentially with lattice depth while interactions
vary weakly with lattice depth \citep{Zwerger}.{]} We choose $J\left(t\right)$
to have the following form:

\begin{equation}
J\left(t\right)=\left(\frac{J_{f}-J_{i}}{2}\right)\tanh\left(\frac{t-t_{c}}{\tau_{Q}}\right)+\left(\frac{J_{f}+J_{i}}{2}\right),\label{eq:hopping amplitude - 1}
\end{equation}

\noindent which corresponds to the experimental scenario of a linear
ramp. Note that $\lim_{t\to-\infty}J\left(t\right)=J_{i}$, and $\lim_{t\to\infty}J\left(t\right)=J_{f}$.
The time scale $\tau_{Q}$ is the characteristic time for $J\left(t\right)$
to cross from $J_{i}$ to $J_{f}$, and $t_{c}$ is the time at which
the middle of the quench is occurring. Other forms of $J\left(t\right)$
which are not linear may lead to differing behaviour in the long-time
limit \citep{Mondal}. For the quench scenario we consider in this
paper, $J_c>J_{f}>J_{i}=0$, where $J_c$
is the critical hopping strength at the superfluid to Mott insulator
phase boundary (for fixed $\mu$).

\section{\label{sec:Numerical results}Numerical Results}

The equations of motion, Eqs.~\eqref{eq:A eqn of motion - 1} and
\eqref{eq:G_K eqn of motion - 1}, form a system of nonlinear Volterra
integral equations that have no known analytical solution, hence we
take a numerical approach to solve them. This presents more of a challenge
than the one-particle-irreducible (1PI) equations of motion obtained
in Ref.~\citep{Kennett} due to the presence of memory kernels that
incorporate the entire history of the system, making explicit the
importance of the quench protocol to the post-quench state. An additional
important feature of the equations of motion is that they are causal,
i.e. all quantities at some later time $t_{f}$ can be obtained by
integration over the known functions for times $t\le t_{f}$. We exploit
this feature of the equations to develop an implicit block-by-block
scheme, closely following Ref.~\citep{Katani}. A detailed discussion
of our numerical scheme is presented in Appendix \ref{sec:Numerical implementation}.

In this section we first 
compare the results of the solutions of  Eqs.~\eqref{eq:A eqn of motion - 1} and
\eqref{eq:G_K eqn of motion - 1} to exact diagonalization (ED) calculations.  
Obtaining acceptable agreement we then present numerical results for the light-cone like
propagation of single-particle spatial correlations in one, two, and
three dimensions for quenches in the Mott insulating regime.

\subsection{Comparison to exact diagonalization calculations}

First, we assess the accuracy of our effective theory by comparing calculations
of the single-particle density matrix $\rho_{1}\left(\Delta\vec{r},t\right)$
obtained from this theory to that from exact diagonalization calculations for small system sizes.
$\rho_{1}\left(\Delta\vec{r},t\right)$ is a natural quantity to study
single-particle spatial correlations, which can be calculated from the equal-time
kinetic Green's function $G_{\vec{k}}^{\left(K\right)}\left(t,t\right)$
as follows:

\begin{align}
\rho_{1}\left(\Delta\vec{r},t\right) & =\frac{1}{N_{\text{sites}}}\sum_{\vec{k}}\cos\left(\vec{k}\cdot\Delta\vec{r}\right)n_{\vec{k}}\left(t\right)\nonumber \\
 & =\frac{1}{2N_{\text{sites}}}\sum_{\vec{k}}\cos\left(\vec{k}\cdot\Delta\vec{r}\right)\left\{ iG_{\vec{k}}^{\left(K\right)}\left(t,t\right)-1\right\} .\label{eq:rho_1 - 1}
\end{align}

\begin{widetext}

\begin{figure}[th]
	\begin{center}
		\includegraphics[scale=0.35]{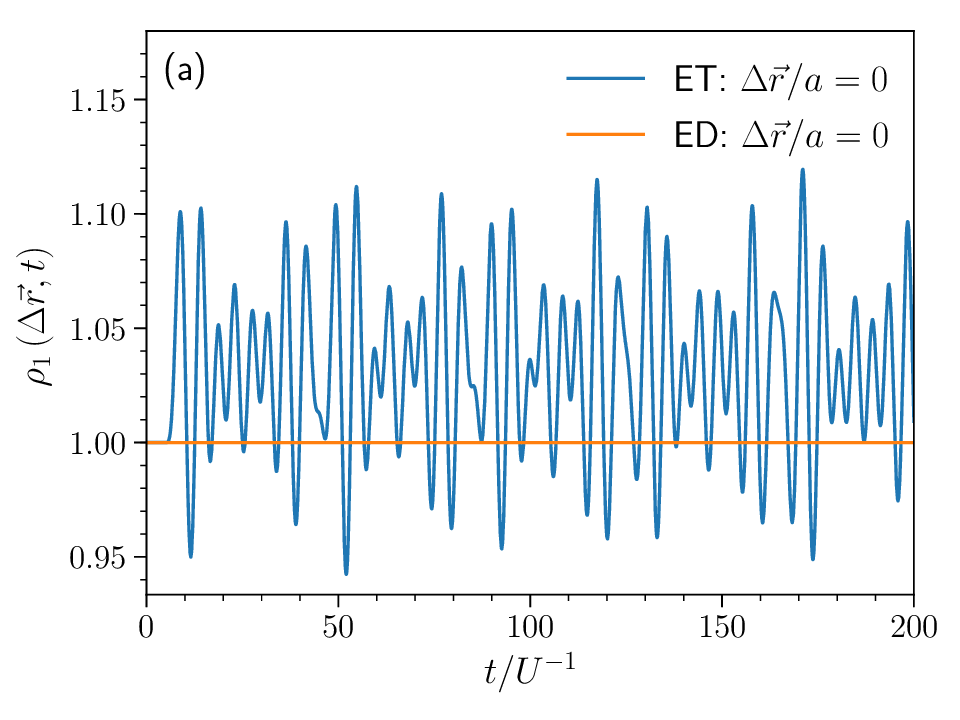}\includegraphics[scale=0.35]{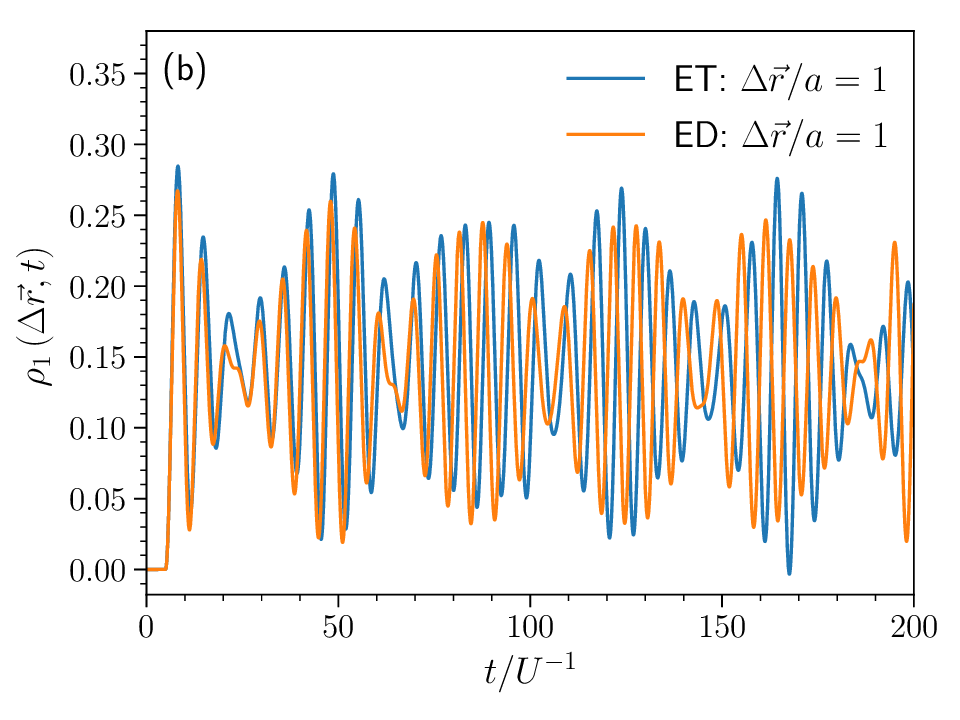}\includegraphics[scale=0.35]{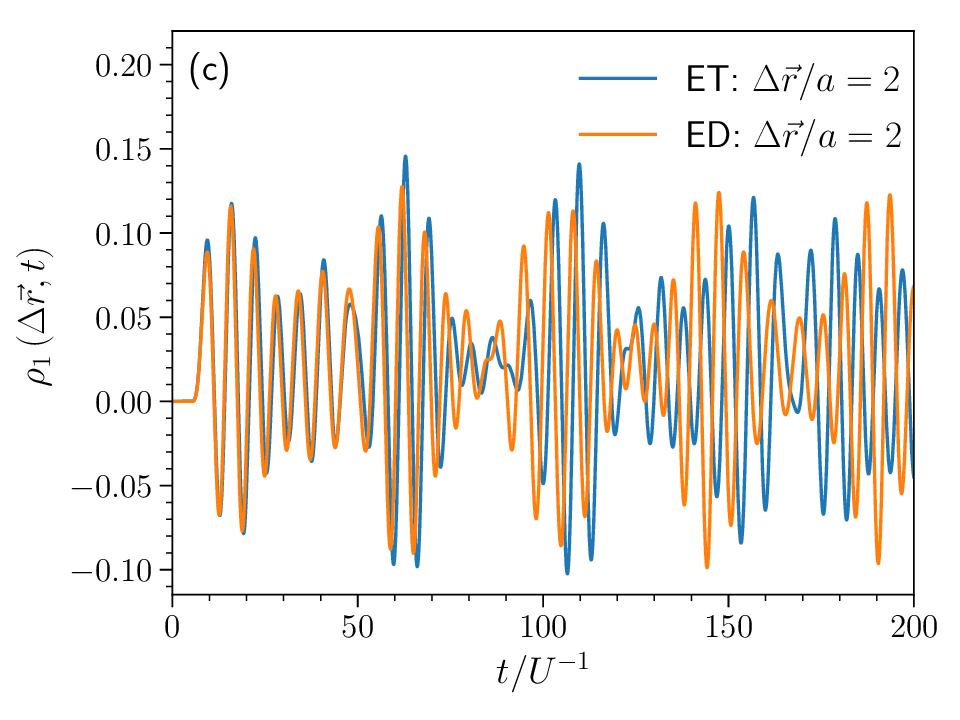}
	\par\end{center}
	\begin{center}
	\includegraphics[scale=0.35]{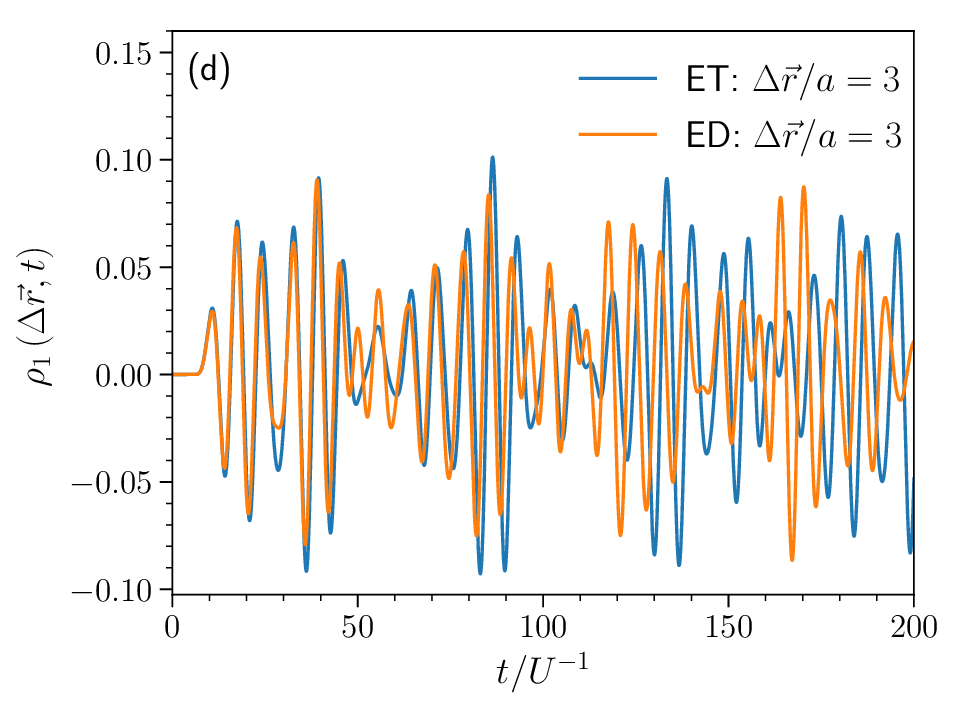}\includegraphics[scale=0.35]{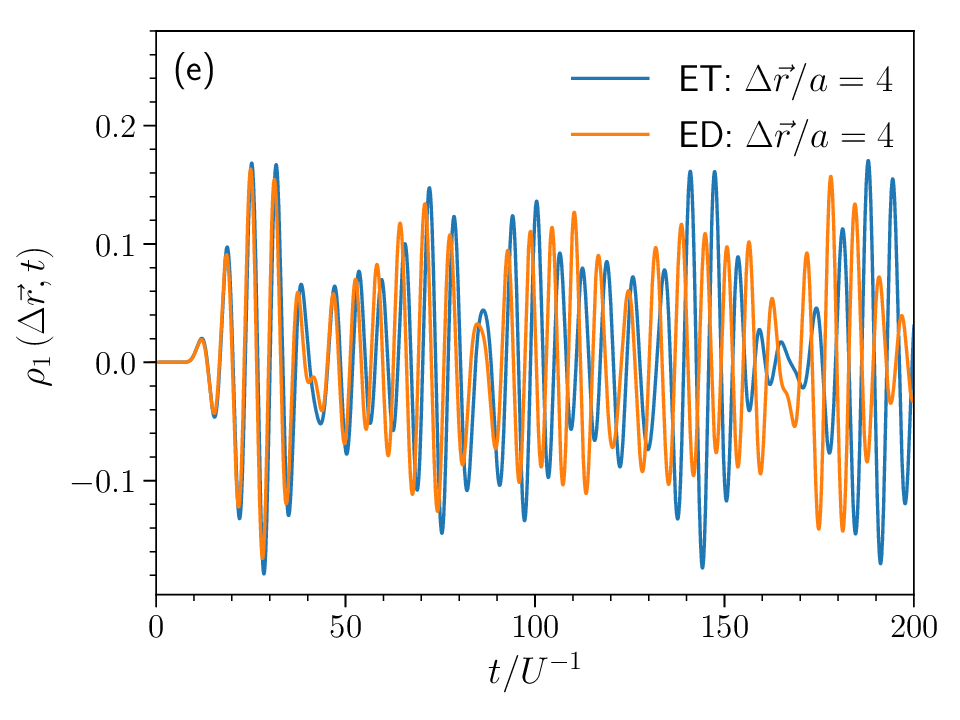}\end{center}
	\caption{(Color online) (a)-(e) Comparison of $\rho_{1}\left(\Delta\vec{r},t\right)$
		        obtained by our ET and by ED for $\Delta\vec{r}/a=0$ to $4$ respectively. The parameters
			        are $\beta U=\infty$, $\mu/U=0.4116$, $J_{f}/U=0.035$, $t_{c}/U^{-1}=5$,
				        $\tau_{Q}/U^{-1}=0.1$, $d=1$, and $N_s=8$. 
				\label{fig:fig2}}
\end{figure}

\begin{figure}[!h]
	\begin{center}
		\includegraphics[scale=0.35]{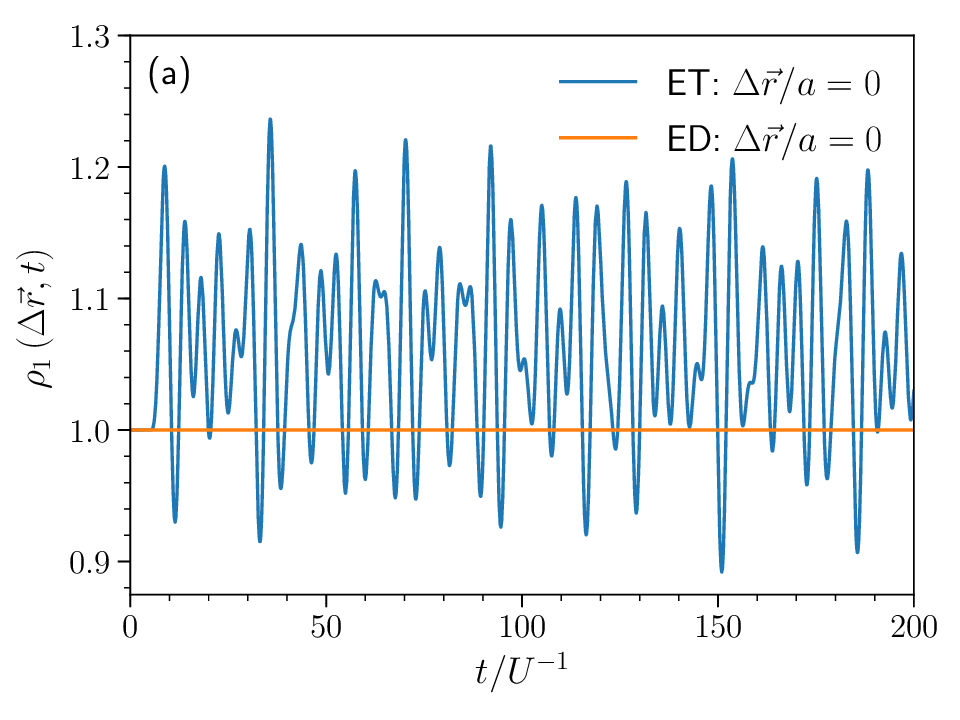}\includegraphics[scale=0.35]{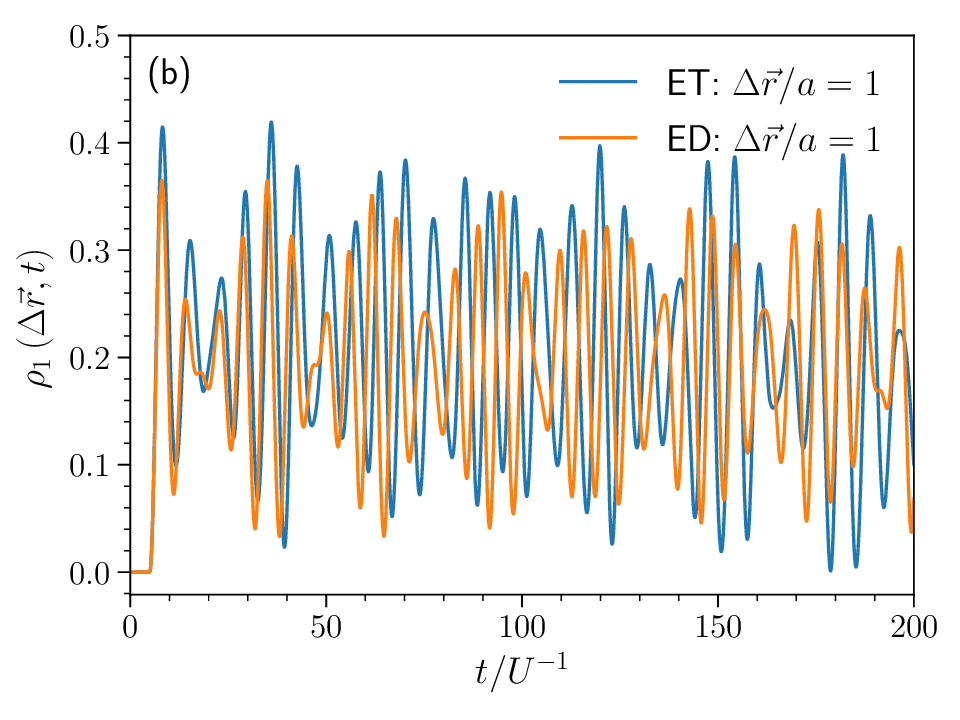}\includegraphics[scale=0.35]{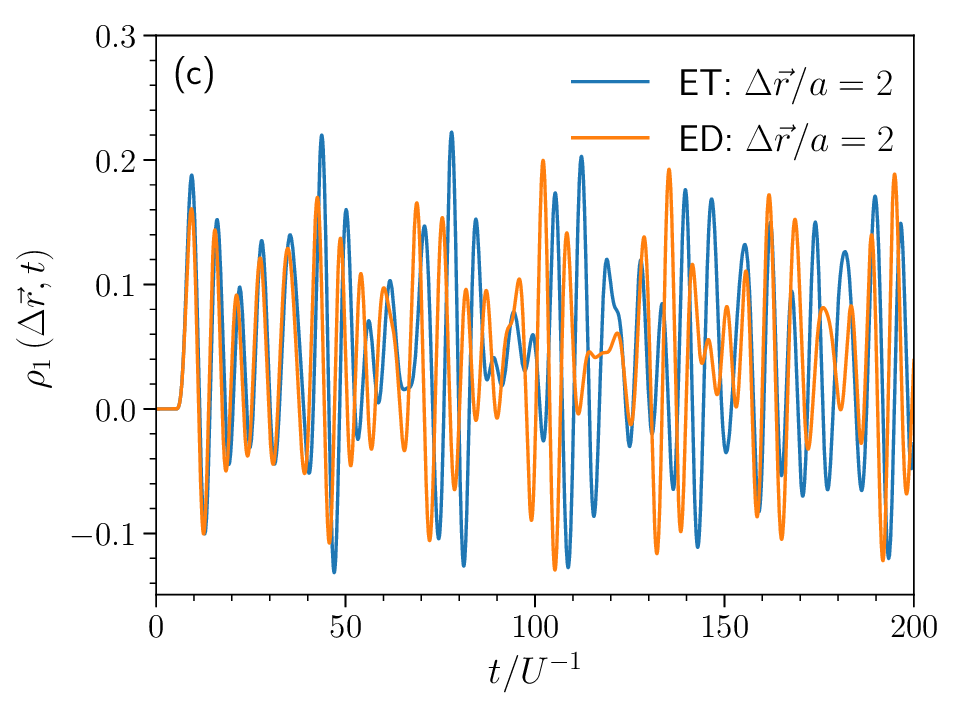}
	\par\end{center}
	\begin{center}
		        \includegraphics[scale=0.35]{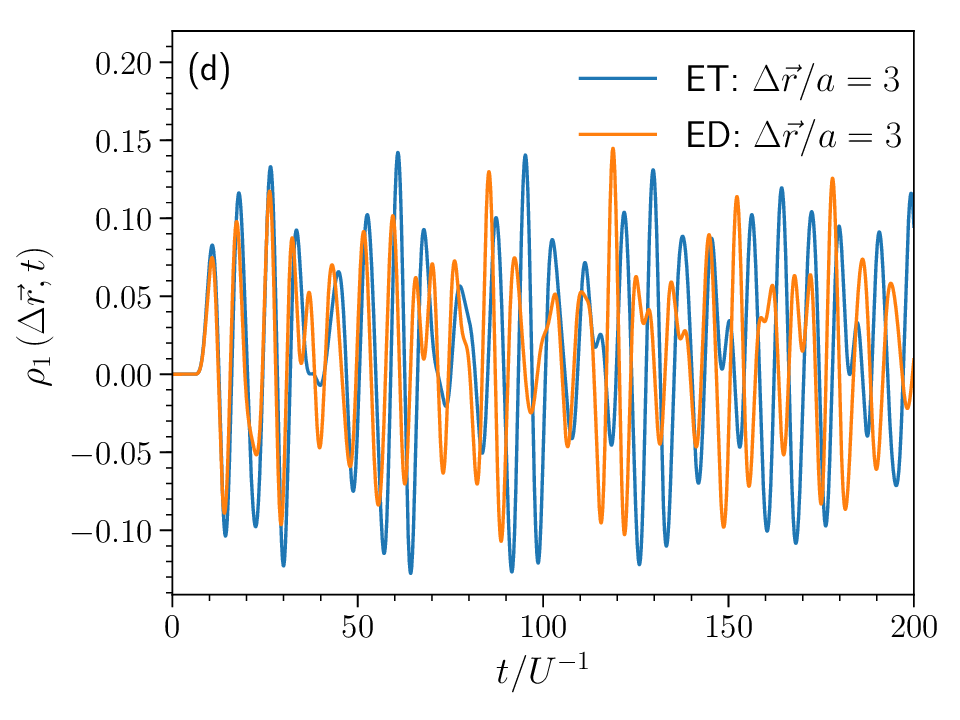}\includegraphics[scale=0.35]{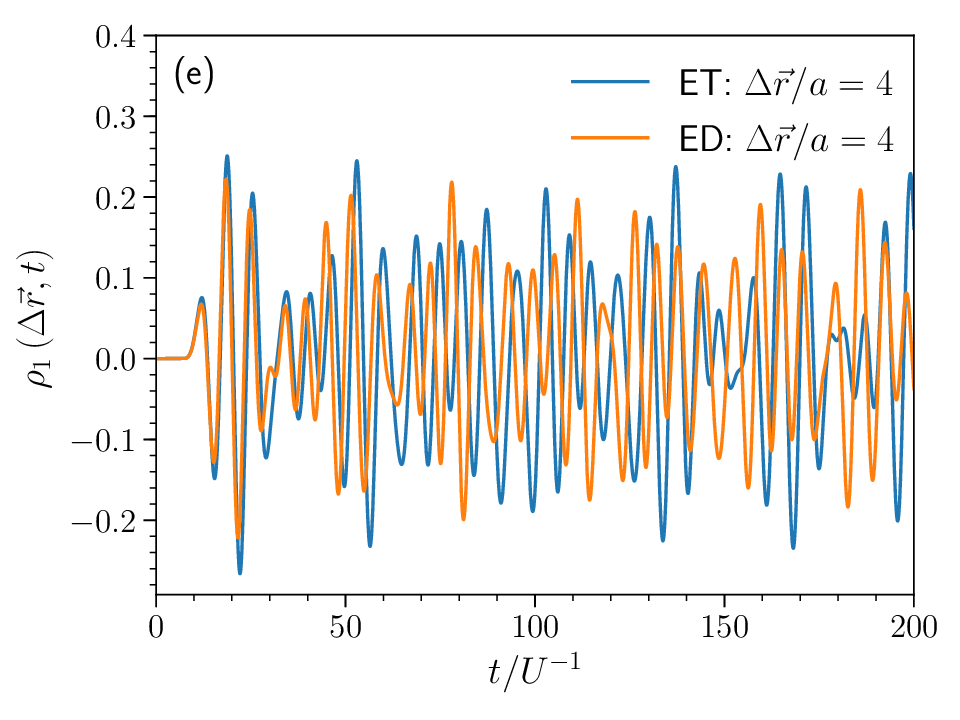}
		\end{center}
		\caption{(Color online) (a)-(e) Comparison of $\rho_{1}\left(\Delta\vec{r},t\right)$
			obtained by our ET and by ED for $\Delta\vec{r}/a=0$ to $4$ respectively. The parameters
			are the same as in Fig.~\ref{fig:fig2} except that $J_{f}/U=0.05$.
		\label{fig:fig3}}
\end{figure}

\begin{figure}[t]
\begin{centering}
\includegraphics[scale=0.52]{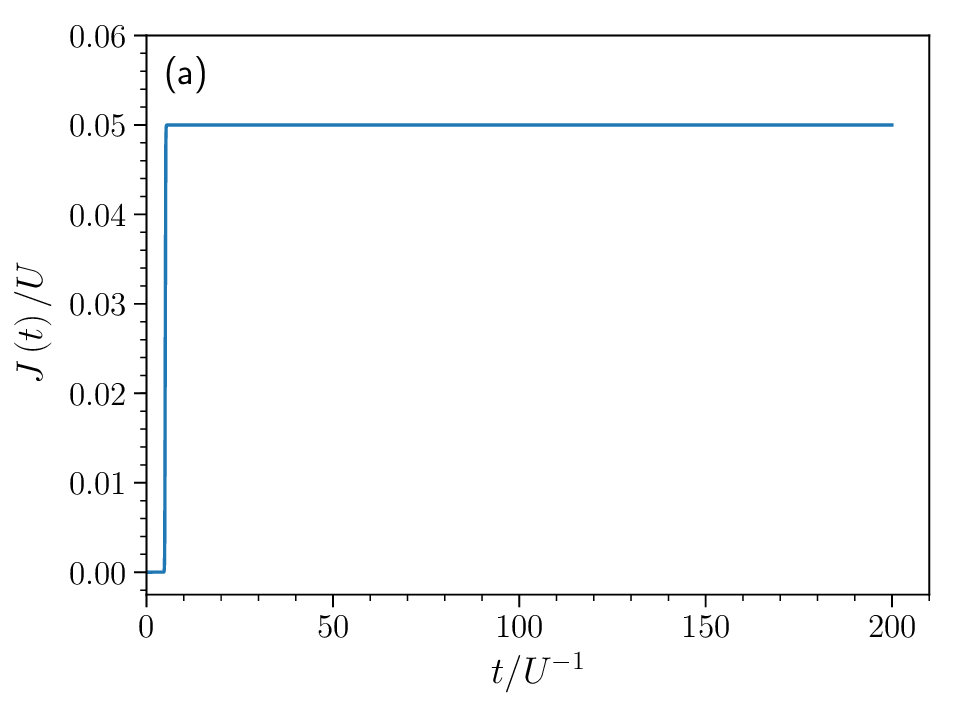}\includegraphics[scale=0.52]{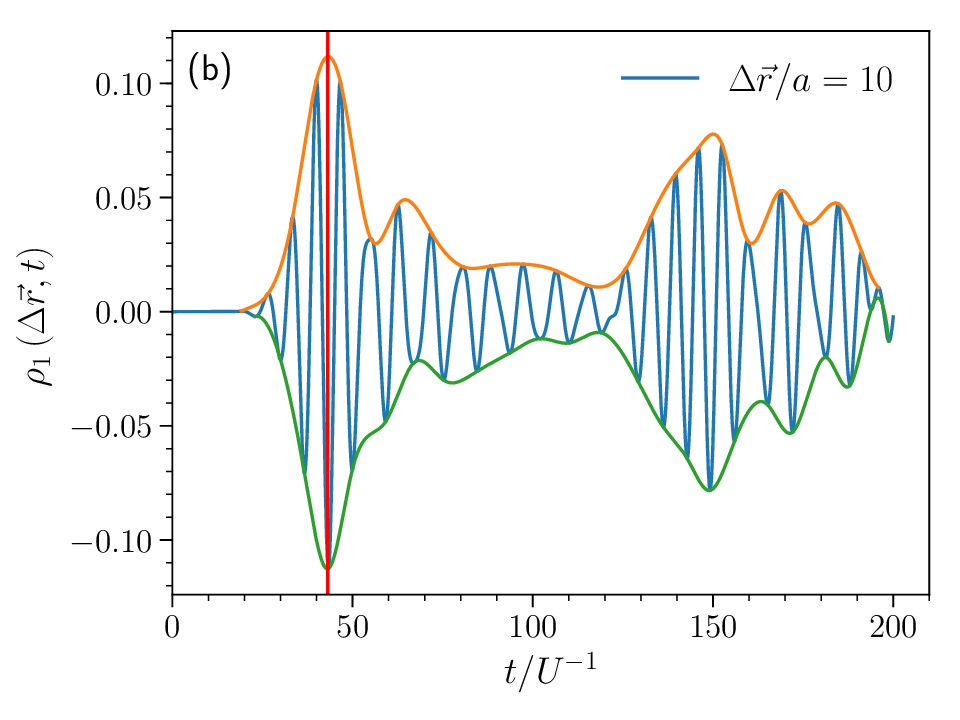}
\par\end{centering}
\begin{centering}
\includegraphics[scale=0.52]{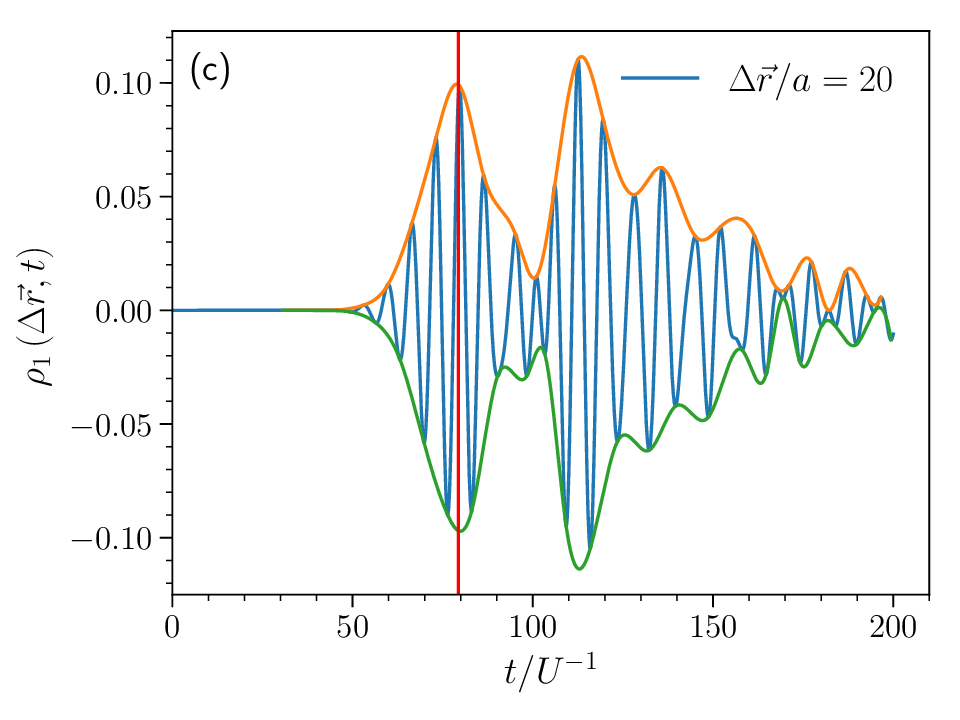}\includegraphics[scale=0.52]{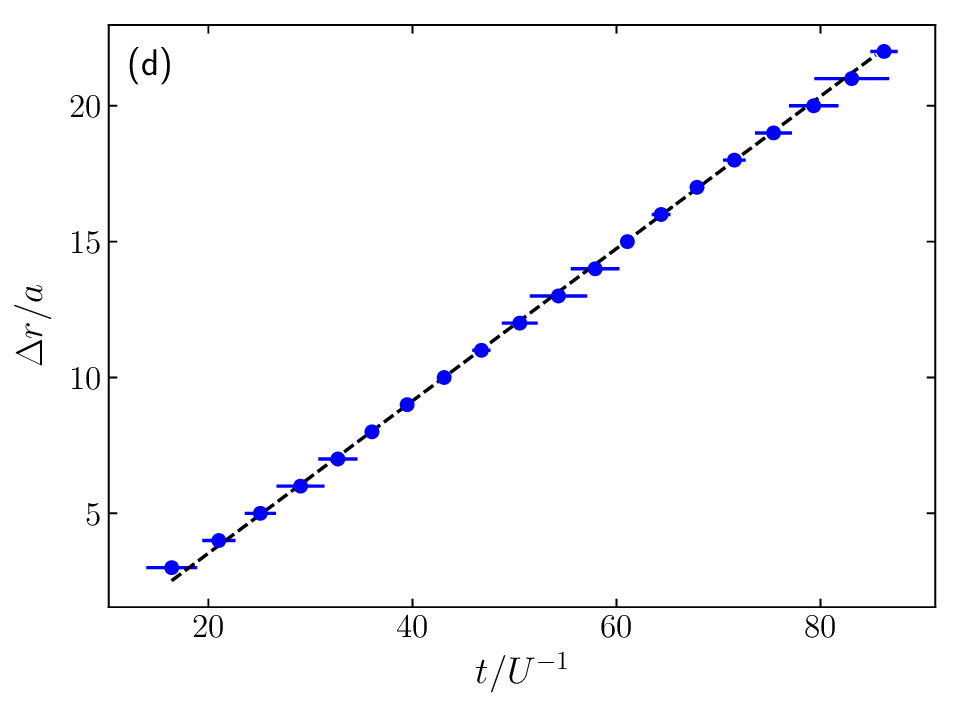}
\par\end{centering}
\caption{(Color online) (a) The evolution of $J\left(t\right) / U$ for quench
  parameters $J_f / U = 0.05$, $t_c / U^{-1} = 5$, and $\tau_Q / U^{-1} = 0.1$;
  (b) dynamics of $\rho_{1}\left(\Delta\vec{r},t\right)$ for
  $\Delta\vec{r} / a = 10$; (c) dynamics of
  $\rho_{1}\left(\Delta\vec{r},t\right)$ for $\Delta\vec{r} / a = 20$;
  (d) scatter plot of the time $t / U^{-1}$ it takes for the
  single-particle correlation front to travel a distance $\Delta r / a$.
  We show a straight line fit to the data. In (b) and (c), the orange
  and green lines trace the envelopes of
  the wavepackets while the red line estimates the position of the centre of the
  first wavepacket. The parameters
  for (b)-(d) include the quench parameters in (a), as well as
  $\mu / U = 0.4116$, $\beta U = 1000$, $d=1$, and $N_s=50$.
\label{fig:fig4}}
\end{figure}

\end{widetext}

In Figs.~\ref{fig:fig2} and \ref{fig:fig3}, we display the time evolution of
$\rho_{1}\left(\Delta\vec{r},t\right)$, obtained from both the effective
theory (ET) and ED, for a quench performed on an 8-site chain ($d=1$; $N_s=8$) with
$\beta = \infty$ ($T = 0$), $\mu / U = 0.4116$,
$t_c / U^{-1}= 5$, and $\tau_Q / U^{-1}= 0.1$. The only differing parameter between
the two figures is the final hopping strength $J_f / U$, where $J_f / U = 0.035$
for Fig.~\ref{fig:fig2} and $J_f / U = 0.05$ for Fig.~\ref{fig:fig3}.

Figure \ref{fig:fig2}(a) plots
$\rho_{1}\left(\Delta\vec{r},t\right)$ for $\Delta\vec{r} / a = 0$,
which is equivalent to the average particle density.
Figure~\ref{fig:fig2}(a) shows that our effective theory leads to small fluctuations in 
the particle number, typically on the order of 5\%. In Appendix~\ref{sec:Particle number conservation}, we 
discuss the origin of these particle number fluctuations.
The results in Figs.~\ref{fig:fig2}(b)-(e) show that this disagreement with ED is confined to 
$\Delta\vec{r} / a = 0$ since for $\Delta\vec{r} / a \neq 0$ 
our method is quantitatively accurate for times up to $ \sim 100 \, U^{-1}$.  At later
times, the beats calculated by our method, begin to become out of phase with those
obtained by ED.

Figures~\ref{fig:fig3}(a)-(e) display the time evolution of $\rho_{1}\left(\Delta\vec{r},t\right)$ for
an identical system to that shown in  Figs.~\ref{fig:fig2}(a)-(e) except that $J_f/U = 0.05$.
For this value of $J_f$, the ET is quantitatively 
accurate for times up to $\sim 50 \, U^{-1}$ when  $\Delta\vec{r} / a \neq 0$.  This is a sufficiently long time window to allow 
the identification of the peak of the first wavepacket in $\rho_{1}\left(\Delta\vec{r},t\right)$ 
at a given $\Delta\vec{r}/a \neq 0$, which we use to determine the velocity at which 
single particle correlations spread.  The good agreement with ED results in 8 site systems gives us 
confidence in the results we obtain in larger systems and higher dimensions where 
comparison with ED is not possible.

\begin{figure*}[th]
\begin{centering}
\includegraphics[scale=0.35]{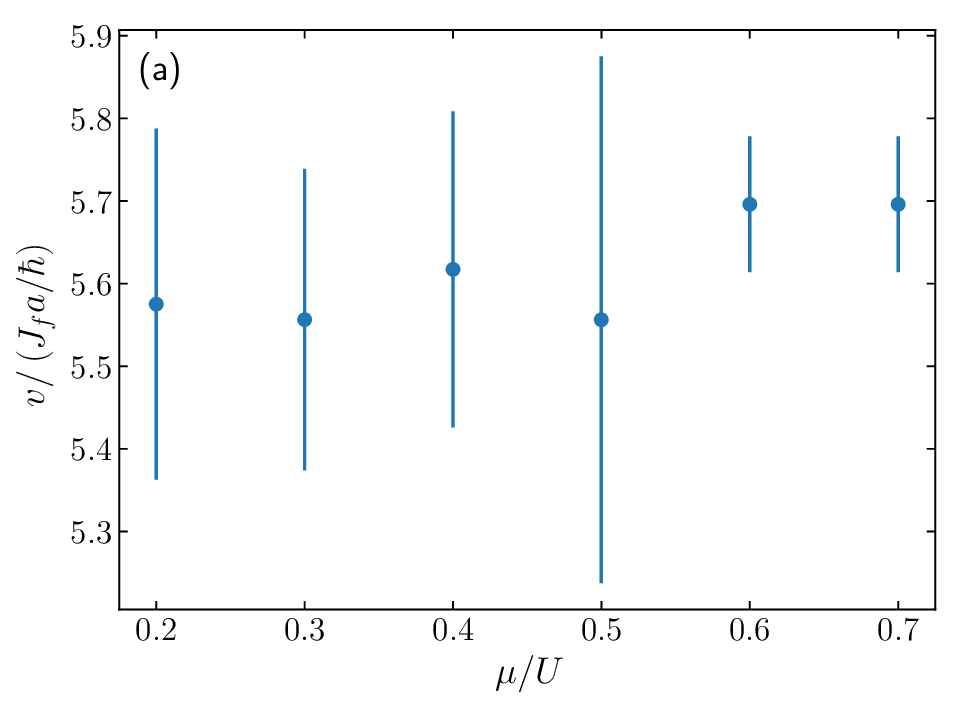}\includegraphics[scale=0.35]{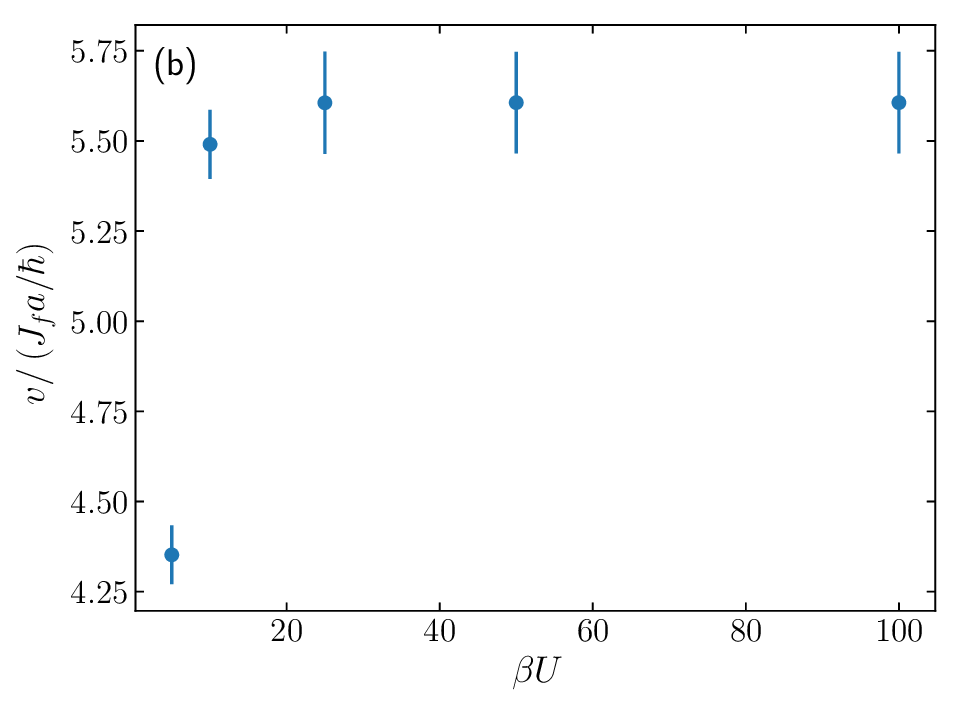}
\includegraphics[scale=0.35]{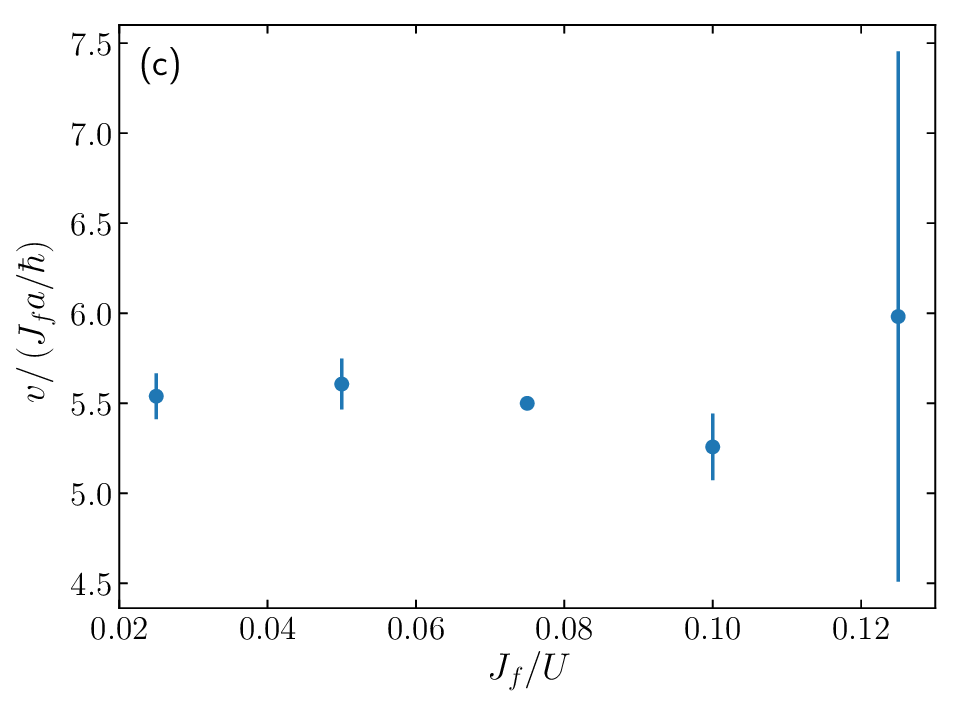}
\par\end{centering}
\caption{(Color online) Scatter plots of the propagation velocity
  $v / \left(J_{f} a / \hbar \right)$ in one dimension as a function of various model
  parameters. In all cases $t_c / U^{-1} = 5$, and $\tau_Q / U^{-1} = 0.1$.
  (a) scatter plot of $v / \left(J_{f} a / \hbar \right)$ as a function of
  $\mu / U$ for a 50 site chain with $\beta U = 1000$, and $J_f / U = 0.05$;
  (b) scatter plot of $v / \left(J_{f} a / \hbar \right)$ as a function of
  $\beta U$ for a 50 site chain with $\mu/U = 0.4116$ and $J_f/U = 0.05$;
  (c) Scatter plot of $v / \left(J_{f} a / \hbar \right)$ as a function of
  $J_f / U$ for a $50$ site chain with $\beta U = 1000$,
  and $\mu / U = 0.4116$.}
\label{fig:fig5}
\end{figure*}

\subsection{\label{subsec:Light-cone spreading of correlations}Light-cone spreading
of single-particle spatial correlations}

In this section, we demonstrate light-cone like spreading \citep{LiebRobinson}
of single particle correlations in one, two and three dimensions, and we compare the
velocities we obtain for the propagation of correlations to existing
results in the field \citep{Bernier,Lauchli,Barmettler,Cheneau,Navez,Natu2,Yanay,Krutitsky}.
We performed calculations of the spreading of correlations in one
(50 site chains), two ($50 \times 50$ systems), and three
dimensions ($28 \times 28 \times 28$ systems) for a variety of different
model parameters and found light-cone like spreading of correlations in all cases.
We present our detailed results below.

\subsubsection{1 dimension}

Before presenting results for the velocity at which single-particle 
correlations spread, we first discuss how we identify this velocity.
In Fig.~\ref{fig:fig4}(b), we display the time evolution of the
single-particle correlation
function $\rho_{1}\left(\Delta\vec{r},t\right)$ for a 50 site chain, with
$\Delta\vec{r} / a = 10$. From this figure, we can see the emergence of
multiple wavepackets after the quench. The orange and green lines
trace the envelopes of these wavepackets
which we determine from an interpolation based on a fourth order spline. 
The red line represents our estimation of the center of the
first wavepacket.
In Fig.~\ref{fig:fig4}(c), where
$\Delta\vec{r} / a = 20$ one can see that the center of the first
wavepacket is shifted to a later
time, i.e. it takes a longer time for the single-particle correlations to
spread out to larger particle separation distances $\Delta r / a$. To track
the propagation of the single-particle correlations, we plot the particle
separation displacement $\Delta\vec{r} / a$ of the first wavepacket
against time $t / U^{-1}$.

We do this for the above 50 site chain system in Fig.~\ref{fig:fig4}(d) and note that the
data is compatible with a linear fit, implying that there is a propagating
front of single-particle correlations that travels through the 1D chain at a
constant velocity $v$. The error bars in Fig.~\ref{fig:fig4}(d) indicate our
uncertainty in determining the centers of the wavepackets. 
Performing a linear fit, we obtain an estimate for the velocity of
$ v=\left(5.6\pm0.1\right)\frac{J_{f}a}{\hbar}$, for this particular set of parameters.

In Fig.~\ref{fig:fig5} we summarize our results for the propagation velocity in one 
dimension as a function of chemical potential, temperature and $J_f/U$ for a 50 site chain.  We see that 
except at temperatures comparable to the melting temperature of the Mott insulator
$\beta U \sim 5$, the velocities we extract all lie in the range 
$5.5 \text{--} 6 \, J_f a / \hbar$ and show little sensitivity to $J_f/U$ or $\mu/U$. These values agree well with the value of 
$v=6 J a / \hbar$ for $\bar{n} = 1$ for the spreading of density-density correlations 
in the limit of infinitely strong interactions 
in 1 dimension obtained by Barmettler \emph{et al. }using a fermionization procedure \cite{Barmettler}.
Experimental data on the spreading of density-density correlations
also lie in the range $5 \text{--} 6 J a / \hbar$ for quenches in the Mott regime \cite{Cheneau}.
In the limit of no interactions Barmettler \emph{et al. }obtained a value of $v=4 J_f a / \hbar$. 
Other recent calculations of the spreading of density density correlations in one dimension found  a value of 
$v=3.7 J a / \hbar$ for weak interactions \cite{Natu2}. Krutitsky \emph{et al. }\cite{Krutitsky} obtained an analytical estimate of $v=3 J_f a / \hbar$ for the single-particle density matrix by
performing a perturbative expansion of the von Neumann equation with respect to the inverse coordination
number, $1 / z$, for small $J_f$.

\subsubsection{2 dimensions}

\begin{figure*}[th]
\begin{centering}
\includegraphics[scale=0.65]{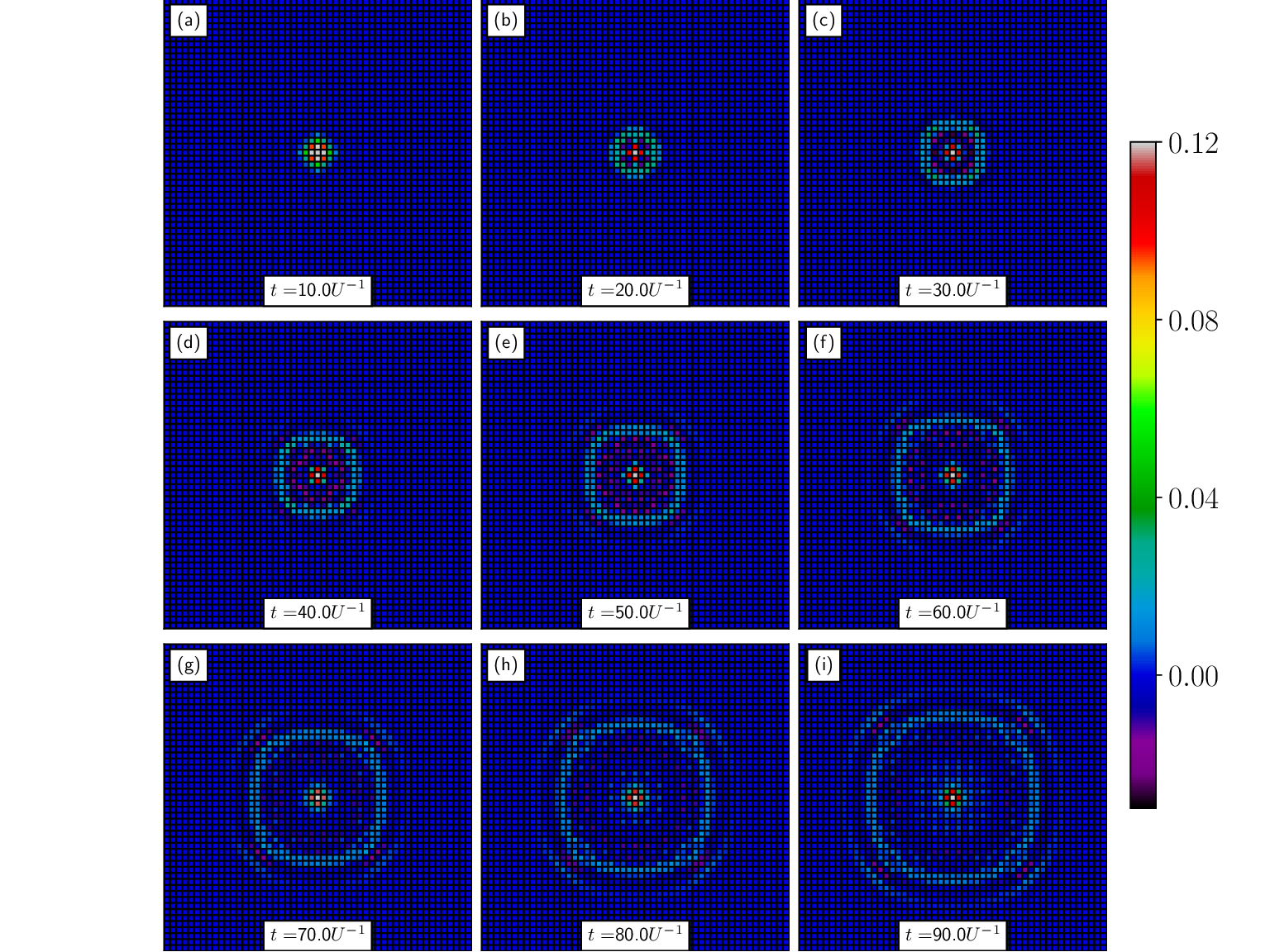}
\par\end{centering}
\caption{(Color online) (a)-(i) Spatial dependency of
  $\rho_{1}\left(\Delta\vec{r},t\right)$
  at different moments in time $t / U^{-1}$ for a $50 \times 50$ site system.
  The parameters are $\beta U=1000$, $\mu/U=0.4136$, $J_{f}/U=0.025$,
  $t_{c}/U^{-1}=5$, and $\tau_{Q}/U^{-1}=0.1$.
\label{fig:fig6}}
\end{figure*}

The spatial dependence of $\rho_{1}\left(\Delta\vec{r},t\right)$ at
different moments in time for a $50 \times 50$ site system is shown in
Fig.~\ref{fig:fig6}, where each pixel represents a different
particle separation displacement $\Delta\vec{r} / a$, and $\Delta\vec{r} / a = 0$
is in the middle of each panel. From the figure, we see that the propagation
of the single-particle correlations is anisotropic, with the propagation
velocity being maximal along the diagonal and minimal along the crystal axes.
Krutitsky \emph{et al. }\cite{Krutitsky} found the same anisotropic spreading
of single-particle correlations for the same quench protocol. Anisotropic
behavior was also observed by Carleo \emph{et al. }\cite{Carleo} in
the spreading of density-density correlations within the superfluid regime.
However, they found that the propagation velocity was maximal along the crystal axes and minimal along
the diagonal, opposite to the behaviour observed here and in Ref.~\cite{Krutitsky} for
the Mott insulator.

We found acquiring estimates for the propagation velocities in higher dimensions
to be somewhat more difficult than in one dimension. This difficulty
is illustrated in Fig.~\ref{fig:fig7} where we extract the propagation
velocities along a crystal axis and the diagonal for the same $50 \times 50$
system considered in Fig.~\ref{fig:fig6}. Figs.~\ref{fig:fig5}(a) and (b)
display the time evolution of $\rho_{1}\left(\Delta\vec{r},t\right)$ for
$\Delta\vec{r} / a = \left(8, 0\right)$ (i.e. along a crystal axis) and
$\Delta\vec{r} / a = \left(8, 8\right)$ (i.e. along a diagonal) respectively.
Upon comparing the two figures, we see that the wavepacket along the
crystal axis is less sharp than that along the diagonal. Consequently, there is
more uncertainy in our estimate of the center of a wavepacket (and hence
the propagation velocity) along a crystal axis than along a diagonal. This
trend extends to three dimensions as well where the wavepackets are
sharpest along the main diagonals, less sharp along the secondary
diagonals, and even less sharp along the crystal axes. The linear fits in
Figs.~\ref{fig:fig7}(c) and (d) yield the following velocity estimates

\begin{align}
  v_{10}=\left(6.8\pm0.3\right)\frac{J_{f}a}{\hbar},\label{eq:v_10 estimate example - 1} \\
  v_{11}=\left(8.1\pm0.1\right)\frac{J_{f}a}{\hbar},\label{eq:v_11 estimate example - 1}
\end{align}

\noindent where $v_{10}$ and $v_{11}$ are the propagation velocities along
the crystal axes and the diagonals respectively. 

Figures~\ref{fig:fig8}(a)-(c) plot the propagation velocities for a $50 \times 50$ system as a function
of $\mu / U$,  $\beta U$, and $J_f / U$ respectively while keeping all the remaining parameters fixed. 
From Figs.~\ref{fig:fig8}(a) and (b),  we see that the propagation velocities are not very sensitive to
$\mu$, or to temperatures below the full melting of the Mott insulating phase ($\beta \gtrsim 5 U$). 
In Fig.~\ref{fig:fig8}(c), we see that there appears to be a slight increase in propagation velocity 
and a decrease in anisotropy for larger $J_{f} / U$.  Extrapolating to larger values of $J_f/U$ it seems
plausible that there might be a value of $J_f/U$ where the spreading of correlations becomes isotropic,
especially given the results of Carleo \emph{et al. }\cite{Carleo} in the superfluid regime, 
where they found the  maximal propagation velocity to be along the crystal axes, not the diagonals.
In future work, we plan
to investigate quench protocols where one crosses the phase boundary into
the superfluid regime which will allow us to  verify if this is indeed the case.
Technically this requires the inclusion of broken symmetry terms in the equations of motion
since these terms are required for a full description of the superfluid regime.

\begin{widetext}

\begin{figure*}[!t]
\begin{centering}
\includegraphics[scale=0.52]{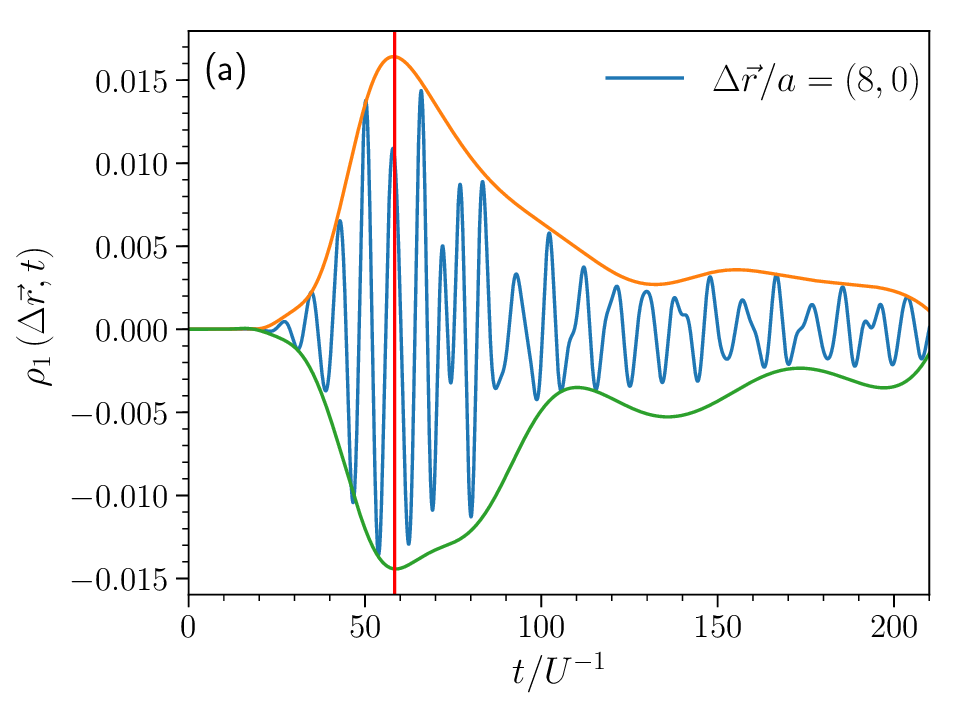}\includegraphics[scale=0.52]{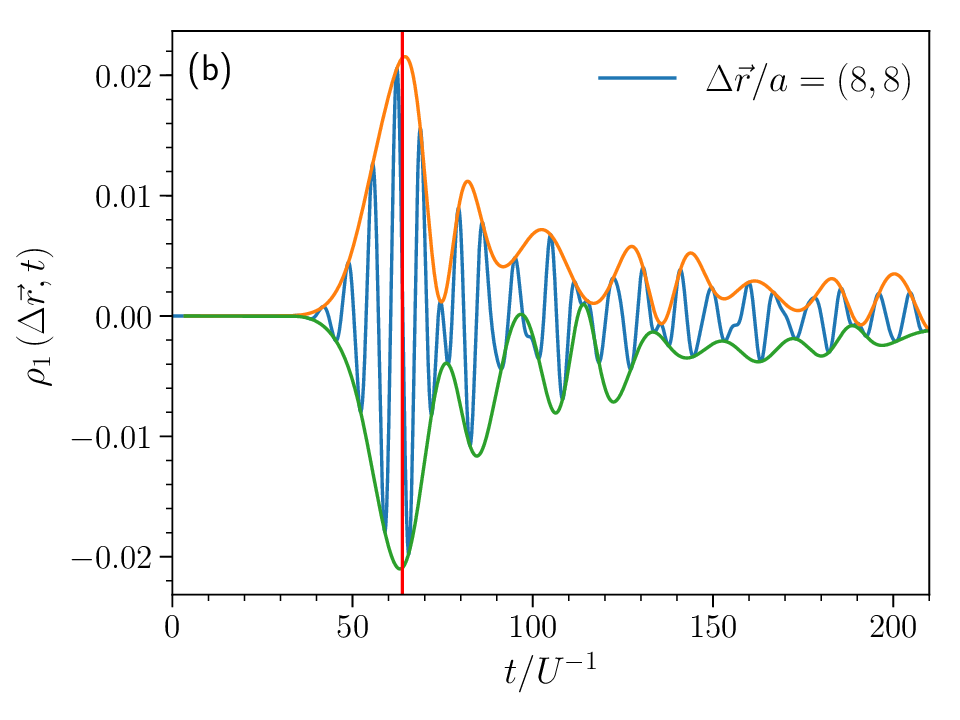}
\par\end{centering}
\begin{centering}
\includegraphics[scale=0.52]{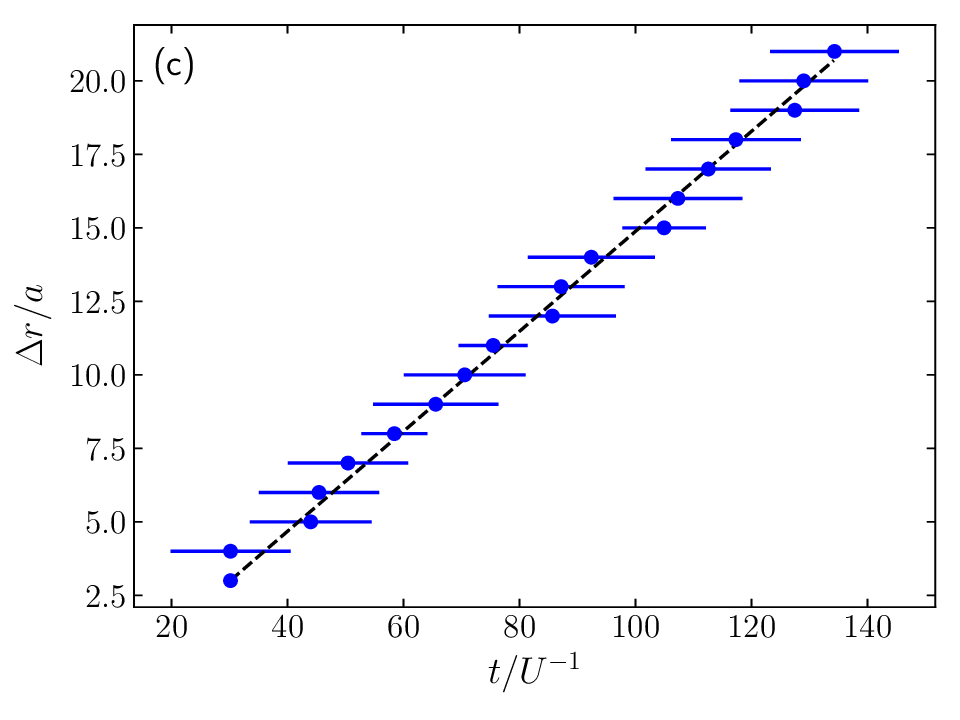}\includegraphics[scale=0.52]{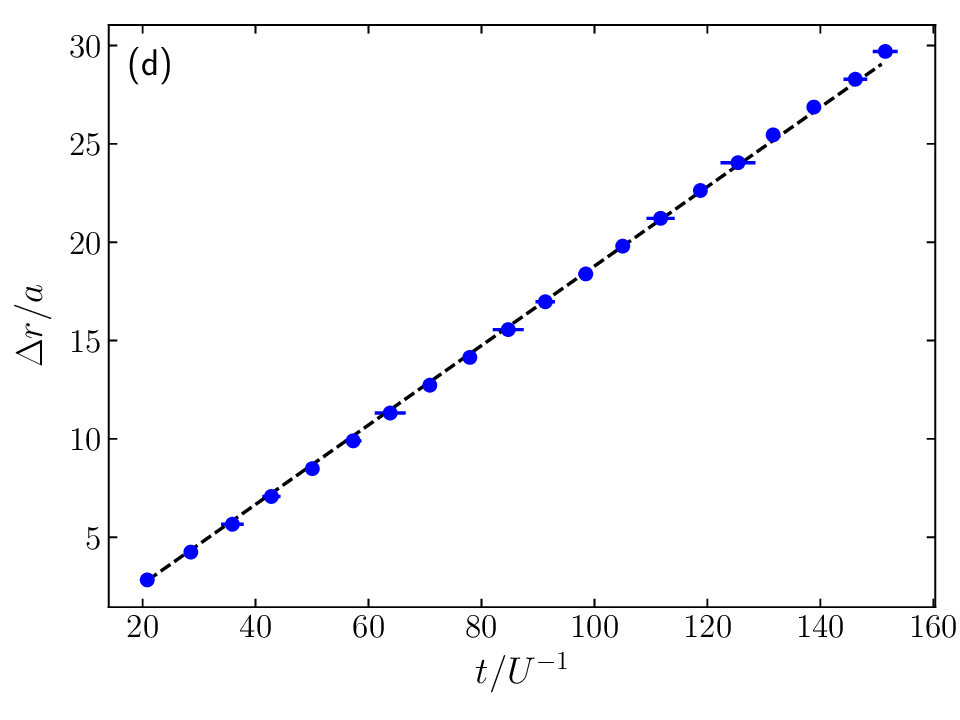}
\par\end{centering}
\caption{(Color online) Tracking the wavefront for a $50 \times 50$ site
  system. (a) Dynamics of $\rho_{1}\left(\Delta\vec{r},t\right)$
  for $\Delta\vec{r} / a = \left(8, 0\right)$; 
  (b) dynamics of $\rho_{1}\left(\Delta\vec{r},t\right)$ for
  $\Delta\vec{r} / a = \left(8, 8\right)$; (c) scatter plot of the time
  $t / U^{-1}$ it takes for the single-particle correlation front to
  travel a distance $\Delta r / a$ along a crystal axis;
  (d) scatter plot of the time $t / U^{-1}$ it takes for the
  single-particle correlation front to travel a distance $\Delta r / a$
  along a diagonal. We show a straight line fit to the data. In (b) and (c),
  the orange and green lines trace the envelopes of
  the wavepackets while the red line estimates the position of the centre of the
  first wavepacket. The parameters for (a)-(d) are
  $\mu / U = 0.4116$, $\beta U = 1000$, $J_f / U = 0.025$,
  $t_c / U^{-1} = 5$, and $\tau_Q / U^{-1} = 0.1$.
\label{fig:fig7}}
\end{figure*}

\begin{figure*}[!h]
\begin{centering}
\includegraphics[scale=0.35]{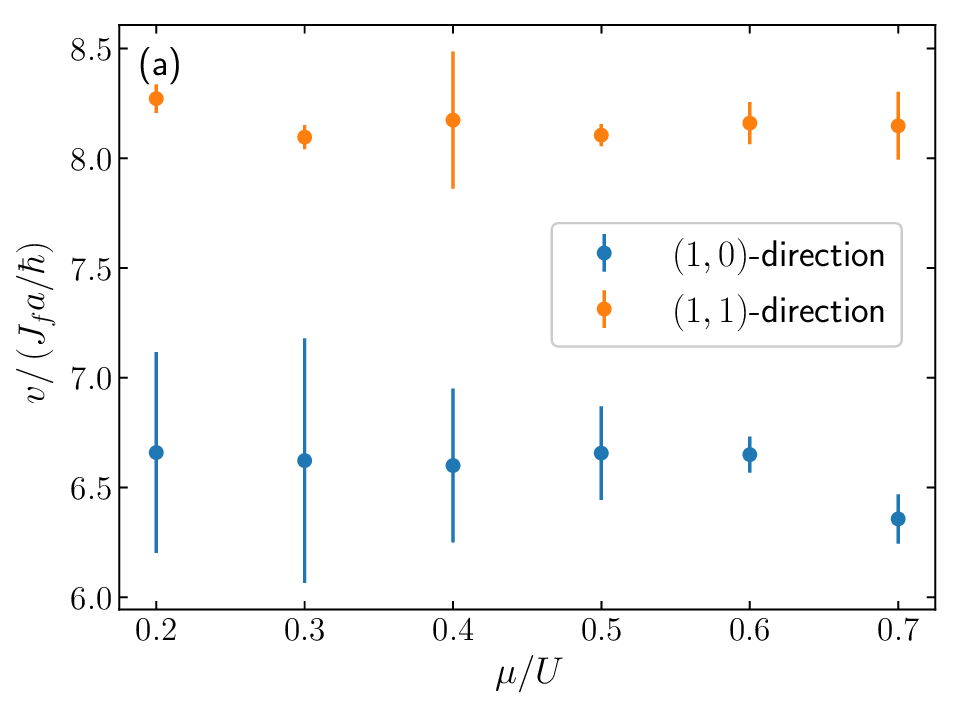}\includegraphics[scale=0.35]{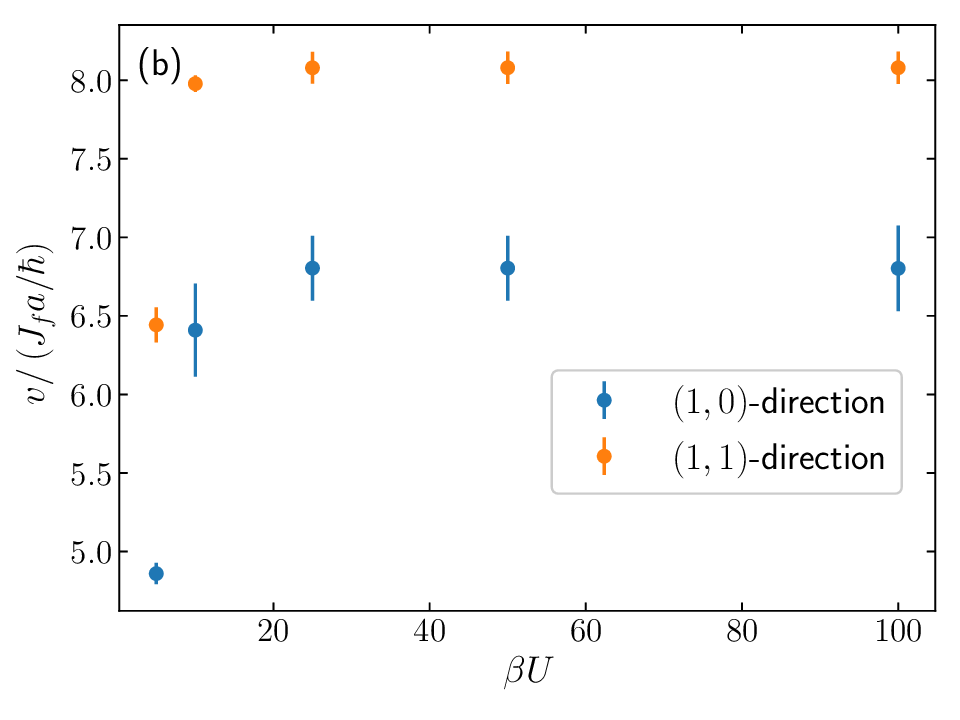}
\includegraphics[scale=0.35]{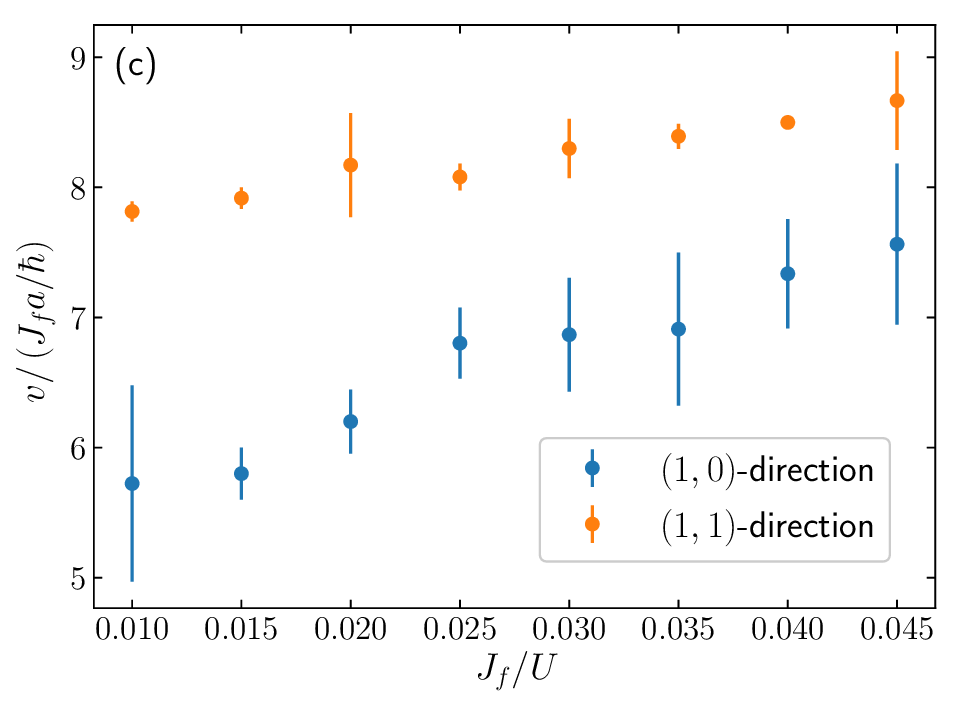}
\par\end{centering}
\caption{(Color online) Scatter plots of the propagation velocity
  $v / \left(J_{f} a / \hbar \right)$ in two dimension for a 50$\times$50
  site system as a function of various model
  parameters. In all cases $t_c / U^{-1} = 5$, and $\tau_Q / U^{-1} = 0.1$.
  (a) scatter plot of $v / \left(J_{f} a / \hbar \right)$ as a function of
  $J_{f} / U$ with $\beta U = 1000$, and $\mu / U = 0.4136$;
  (b) scatter plot of $v / \left(J_{f} a / \hbar \right)$ as a function of
  $\beta U$ with $\mu/U = 0.4136$ and $J_f/U = 0.025$;
  (c) Scatter plot of $v / \left(J_{f} a / \hbar \right)$ as a function of
  $J_{f} / U$ with $\beta U = 1000$,
  and $\mu / U = 0.4136$.}
\label{fig:fig8}
\end{figure*}

\begin{figure*}[!t]
\begin{centering}
\includegraphics[scale=0.35]{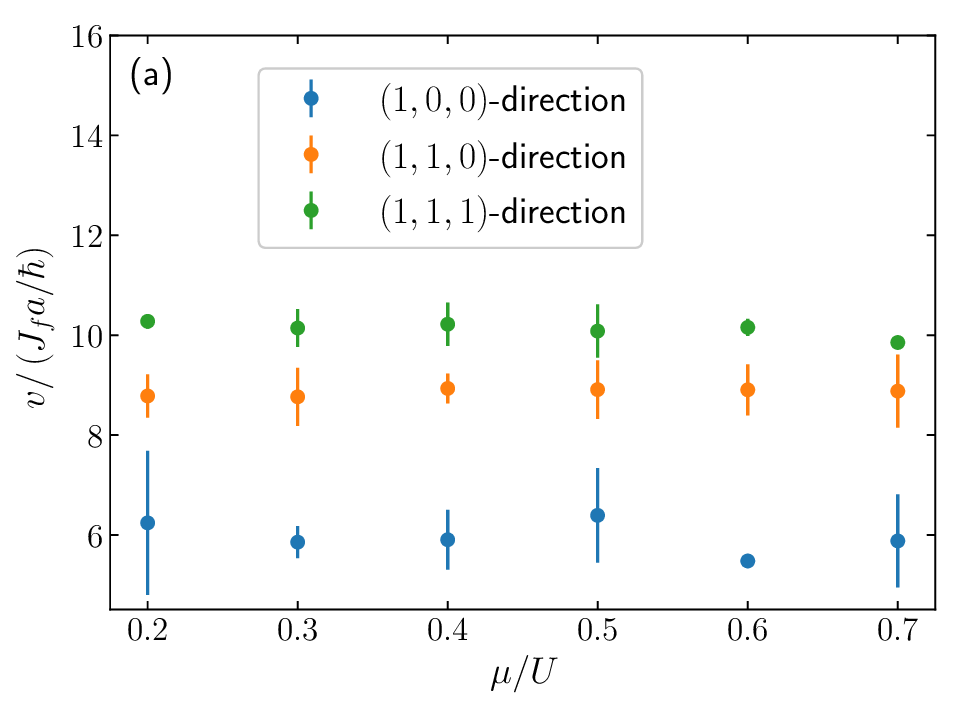}\includegraphics[scale=0.35]{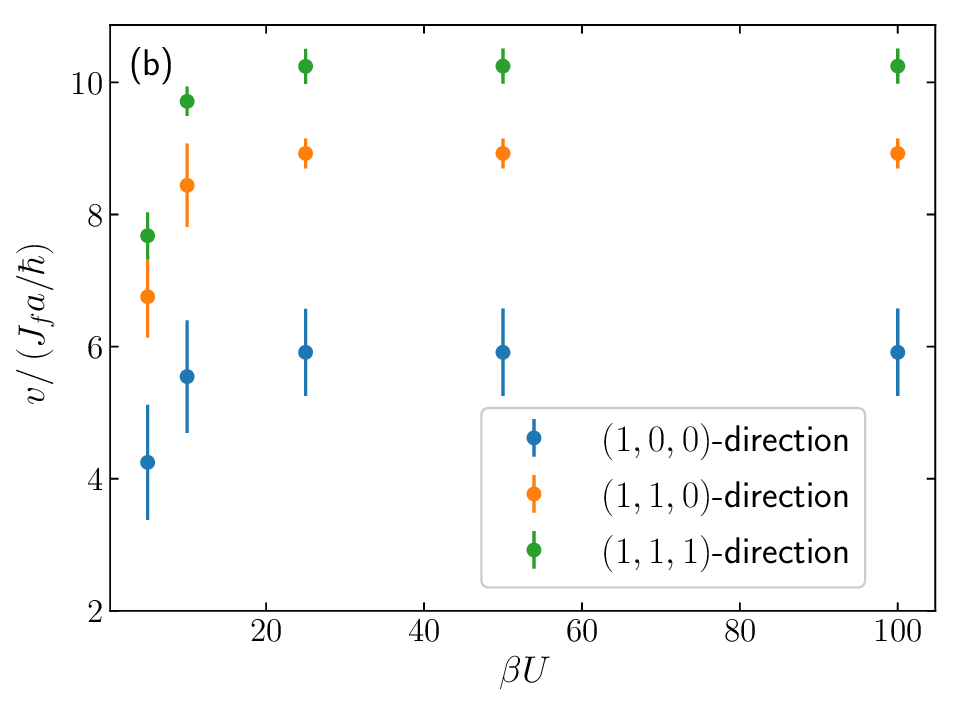}
\includegraphics[scale=0.35]{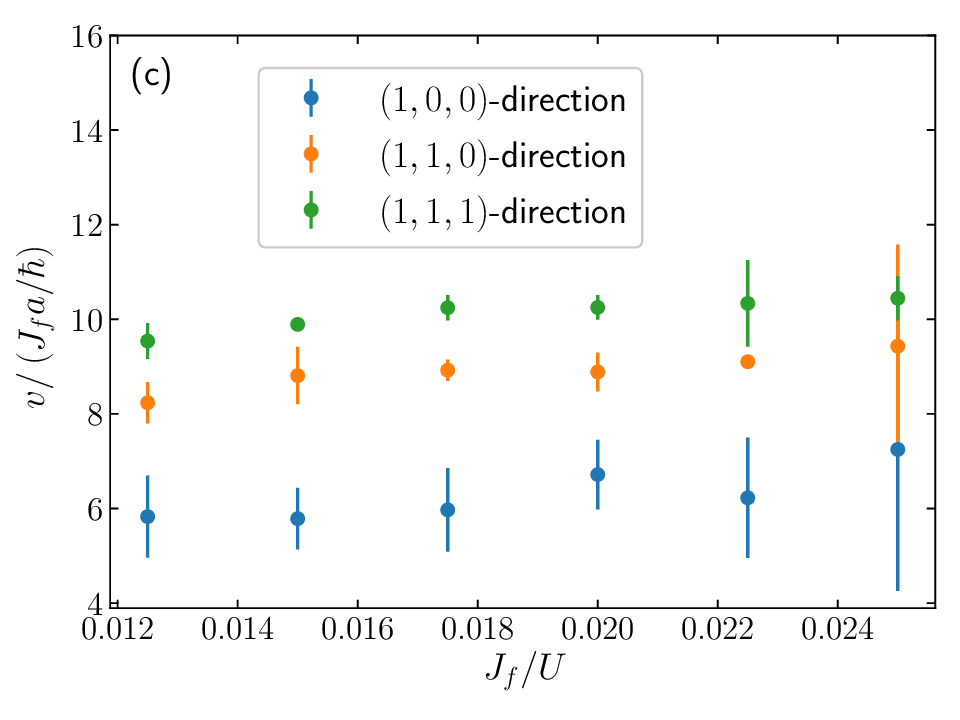}
\par\end{centering}
\caption{(Color online) Scatter plots of the propagation velocity
  $v / \left(J_{f} a / \hbar \right)$ in three dimension for a 28$\times$28
  site system as a function of various model
  parameters. In all cases $t_c / U^{-1} = 5$, and $\tau_Q / U^{-1} = 0.1$.
  (a) scatter plot of $v / \left(J_{f} a / \hbar \right)$ as a function of
  $J_{f} / U$ with $\beta U = 1000$, and $\mu / U = 0.4132$;
  (b) scatter plot of $v / \left(J_{f} a / \hbar \right)$ as a function of
  $\beta U$ with $\mu/U = 0.4132$ and $J_f/U = 0.0175$;
  (c) Scatter plot of $v / \left(J_{f} a / \hbar \right)$ as a function of
  $J_{f} / U$ with $\beta U = 1000$,
  and $\mu / U = 0.4132$.}
\label{fig:fig9}
\end{figure*}

\end{widetext}

In 2
dimensions, the velocities $v_{10}$ we obtained along the crystal axes ranged from
$5.7\text{--}7.6 \,J a / \hbar$ whereas the velocities $v_{11}$ along the diagonal
ranged from $7.8\text{--}8.7 \, J a / \hbar$. The only other related study
that we are aware of is that of Krutitsky \emph{et al. }\cite{Krutitsky},
where they obtained analytical estimates of $v_{10}=3 J a / \hbar$ and
$v_{11}=3 \sqrt{2} J a / \hbar$ for the crystal axes and diagonals respectivekly. 
It is worth pointing out that Krutitsky \emph{et al.} also performed 
numerical calculations of the single-particle
correlation spreading beyond their lowest order analytical calculations,
however they did not report any velocity estimates based on their numerical
data.  One prediction of Krutitsky {\it et al.} that does seem reasonably 
robust is the ratio $v_{11}/v_{10}$, for which their lowest order estimate is $\sqrt{2}$.
Examination of Fig.~\ref{fig:fig8}(c) shows that our results are consistent with 
$v_{11}/v_{10} \simeq \sqrt{2}$ for small $J_f/U$, with the ratio decreasing with increasing
$J_f/U$.

\subsubsection{3 dimensions}

We see similar behaviour in three dimensions compared to that in two dimensions, 
as displayed in Fig.~\ref{fig:fig9} where we see that the velocity depends 
strongly on crystal direction but is otherwise relatively insensitive to changes in
chemical potential, temperature or final hopping value $J_f$.  The trend towards
increasing isotropy in the spread of correlations as $J_f/U$ increases is much less 
pronounced than in two dimensions, perhaps because we consider smaller values of 
$J_f$ than in two dimensions.  To the best of our knowledge, our work is the 
first to calculate propagation velocities for correlations in three dimensions 
for the BHM, and we find that $v_{100} \sim 6 \, J_f a/\hbar$, $v_{110} 
\sim 8.5 \, J_f a/\hbar$ and $v_{111} \sim 10 \, J_f a/\hbar$.

\section{\label{sec:Discussion and conclusions}Discussion and conclusions}

The ability to address single sites in cold atom experiments \citep{Bakr}
has allowed for experimental exploration of spatio-temporal correlations
in the BHM \citep{Cheneau}. This has led to theoretical investigations
of these correlations in both one \citep{Barmettler} and higher dimensions
\citep{Carleo,Natu2,Yanay,Krutitsky} in the presence of a quench.
In dimensions higher than one, where numerical approaches are limited,
a theoretical challenge has been to develop a framework which can
treat correlations in both the superfluid and Mott insulating phases
over the course of a quench. In a previous paper \citep{Fitzpatrick},
we developed a formalism that allows for such a description of the space
and time dependence of single-particle correlations.  The specific 
approach we took was to derive a 2PI effective action for the BHM 
using the KP contour, building on the 1PI real-time strong-coupling low-energy theory developed
in Ref.~\citep{Kennett} which generalized the imaginary-time theory
developed in Ref.~\citep{SenguptaDupuis}. From this 2PI effective
action we were able to derive equations of motion that treat the superfluid
order parameter and the full two-point Green's functions
on equal footing. One of the attractive features of the formalism
is that it is applicable even in the limit of low occupation number
per site. 

Here, we used the formalism to study out of equilibrium dynamics, 
focusing on the light-cone like spreading of single-particle correlations
after a quench.
We considered quenches in the Mott insulator phase and 
solved the equations of motion for the single-particle density matrix $\rho_{1}\left(\Delta\vec{r},t\right)$.
From the calculation of $\rho_{1}\left(\Delta\vec{r},t\right)$,
we demonstrated light-cone like spreading
of single-particle correlations in one, two and three dimensions. The range of propagation velocities that we obtain
in one dimension over the range of parameter values we consider
agree well with recent theoretical \cite{Barmettler}
and experimental results \cite{Cheneau}. Interestingly, it seems that
the results we obtain for single-particle correlations appear to be 
similar to those obtained for density-density correlations.
In higher dimensions, we find
that there is an anisotropic spreading of correlations, where the
propagation velocity is maximal along the main diagonal and minimal
along crystal axes. Similar anisotropic spreading of correlations
was observed in Ref.~\cite{Krutitsky}.  We also observed that at least
in two dimensions, the degree of anisotropy appears to  diminish 
with increasing final hopping strength $J_f$. This raises the question 
of whether the spreading becomes isotropic for $J_f$ in the vicinity of $J_c$,
particularly given that there has been the prediction that in the 
superfluid regime the propagation velocity is maximal along the crystal axes,
rather than the diagonals \cite{Carleo}.  To address these questions within 
our formalism requires a more careful treatment
of the equations of motion. One needs to include broken symmetry terms which
become relevant upon entering the superfluid regime. We defer this task to
future work.

The space and time dependence of correlations after a quantum quench
give insight into the propagation of excitations generated by that
quench, and hence we hope that the formalism we have developed here
will allow further theoretical investigation of the excitations after
quenches in the BHM, to complement experimental efforts in the same
direction. In future work we plan to investigate a broader range of
quench protocols, and generalizations such as the inclusion of a harmonic
trap, coupling to a bath \citep{Robertson,Dalidovich,Guo}, disorder \citep{White,Pasienski,Choi,Meldgin},
or multicomponent \citep{Rubio-Abadal} Bose Hubbard models.
\begin{acknowledgments}
This work was supported by NSERC. 
\end{acknowledgments}

\appendix

\section{\label{sec:Numerical implementation}Numerical implementation of
solution of equations of motion}

In this Appendix, we describe in more detail the numerical implementation
of the solutions to the equations of motion. We begin by rewriting
Eqs.~\eqref{eq:A eqn of motion - 1} and \eqref{eq:G_K eqn of motion - 1}
in a slightly more compact form:

\begin{align}
A_{\vec{k}}\left(t,t^{\prime}\right) & =\mathcal{A}\left(t-t^{\prime}\right)\nonumber \\
 & \quad+\int_{t^{\prime}}^{t}dt^{\prime\prime}K_{\vec{k}}^{\left(1\right)}\left(t,t^{\prime\prime},n\left(t^{\prime\prime}\right)\right)A_{\vec{k}}\left(t^{\prime\prime},t^{\prime}\right),\label{eq:A eqn of motion - 2}\\
G_{\vec{k}}^{\left(K\right)}\left(t,t^{\prime}\right) & =\mathcal{G}^{\left(K\right)}\left(t-t^{\prime}\right)\nonumber \\
 & \quad+\int_{0}^{t}dt^{\prime\prime}K_{\vec{k}}^{\left(1\right)}\left(t,t^{\prime\prime},n\left(t^{\prime\prime}\right)\right)G_{\vec{k}}^{\left(K\right)}\left(t^{\prime\prime},t^{\prime}\right)\nonumber \\
 & \quad+\int_{0}^{t^{\prime}}dt^{\prime\prime}K_{\vec{k}}^{\left(2\right)}\left(t,t^{\prime\prime},n\left(t^{\prime\prime}\right)\right)A_{\vec{k}}\left(t^{\prime\prime},t^{\prime}\right),\label{eq:G_K eqn of motion - 2}
\end{align}

\noindent where we define the kernels

\begin{align}
K_{\vec{k}}^{\left(1\right)}\left(t,t^{\prime\prime},n\left(t^{\prime\prime}\right)\right) & =-i\mathcal{A}\left(t-t^{\prime\prime}\right)\Sigma_{\vec{k}}^{\left(HFB\right)}\left(t^{\prime\prime}\right),\label{eq:K^(1) - 1}\\
K_{\vec{k}}^{\left(2\right)}\left(t,t^{\prime\prime},n\left(t^{\prime\prime}\right)\right) & =i\mathcal{G}^{\left(K\right)}\left(t-t^{\prime\prime}\right)\Sigma_{\vec{k}}^{\left(HFB\right)}\left(t^{\prime\prime}\right).\label{eq:K^(2) - 1}
\end{align}

\noindent We include $n\left(t^{\prime\prime}\right)$ in the kernel
arguments to emphasize the fact that both kernels are functions of
the particle density. The presence of $n\left(t^{\prime\prime}\right)$
in the kernels couples the equations of motion for fixed quasi-momentum
$\vec{k}$ to the remaining equations (with different $\vec{k}$)
since $n\left(t^{\prime\prime}\right)$ is calculated from $\sum_{\vec{k}}G_{\vec{k}}^{\left(K\right)}\left(t^{\prime\prime},t^{\prime\prime}\right)$.
Moreover, for $t\ge t^{\prime}$, the calculation of $A_{\vec{k}}\left(t,t^{\prime}\right)$
and $G_{\vec{k}}^{\left(K\right)}\left(t,t^{\prime}\right)$ depends
on $n\left(t\right)$, not simply the history. These nonlinearities
complicate the numerical solution as we must resort to implicit methods.
At a general level, the simplest method to solve such a nonlinear
system is to apply a self-consistent approach, which we do in this
paper. For each timestep in $t$, we start by guessing the value of
$n\left(t\right)$, then we solve each equation separately for values of 
$t^\prime$ in the range $t\ge t^{\prime}\ge0$
using an explicit numerical approach, then we use our calculation
of the $G_{\vec{k}}^{\left(K\right)}\left(t,t\right)$'s to update
$n\left(t\right)$, and then we repeat until we obtain convergence. Once convergence
is achieved, we take another timestep in $t$, then repeating the
above procedure starting with $t^{\prime}=0$ to $t^{\prime}=t$.
One can guess $n\left(t\right)$ using the final value for $n\left(t-\Delta t\right)$
or by doing an extrapolation based on several previous timesteps.

After guessing/updating the value of $n\left(t\right)$, we implement
a modified block-by-block algorithm based on that in Ref.~\citep{Katani}.
The block-by-block method uses a combination of Simpson's rule and
Lagrange interpolation points to discretize the equations of motion
in such a way to generate a system of equations in terms of multiple
unknowns that can then be solved simultaneously. For example, if
we introduce the following discretization notation

\begin{equation}
F_{m}=F\left(m\Delta t\right),\label{eq:discretization notation - 1}
\end{equation}

\noindent then for fixed $m\ge m^{\prime}$, after applying the block-by-block
procedure, we obtain a pair of simultaneous equations for $\left[A_{\vec{k}}\right]_{2m+1,2m^{\prime}}$
and $\left[A_{\vec{k}}\right]_{2m+2,2m^{\prime}}$, a single equation
for $\left[A_{\vec{k}}\right]_{2m+1,2m^{\prime}+1}$, $\left[A_{\vec{k}}\right]_{2m+2,2m^{\prime}+1}$
and $\left[A_{\vec{k}}\right]_{2m+2,2m^{\prime}+2}$ each, a pair
of simultaneous equations for $\left[G_{\vec{k}}^{\left(K\right)}\right]_{2m+1,2m^{\prime}}$
and $\left[G_{\vec{k}}^{\left(K\right)}\right]_{2m+2,2m^{\prime}}$,
a pair of simultaneous equations for $\left[G_{\vec{k}}^{\left(K\right)}\right]_{2m+1,2m^{\prime}+1}$
and $\left[G_{\vec{k}}^{\left(K\right)}\right]_{2m+2,2m^{\prime}+1}$,
and finally a single equation for $\left[G_{\vec{k}}^{\left(K\right)}\right]_{2m+2,2m^{\prime}+2}$.
These ``block'' equations should be solved in the order as is written
above since each block equation depends on the solutions to the block
equations previous to it.

In summary, our numerical solution can be outlined as follows:
\begin{enumerate}
\item Set $m=0$.
\item \label{enu:block-by-block alg step 2}Guess values for $n_{2m+1}$
and $n_{2m+2}$.
\item \label{enu:block-by-block alg step 3}For each $\vec{k}$:
\begin{enumerate}
\item[] For $m^{\prime}=0,\ldots,m$:
Solve block equations.
\end{enumerate}
\item Update $n_{2m+1}$ and $n_{2m+2}$ from the new 
$\left[G_{\vec{k}}^{\left(K\right)}\right]_{2m+1,2m^{\prime}+1}$
and $\left[G_{\vec{k}}^{\left(K\right)}\right]_{2m+2,2m^{\prime}+2}$
using Eqs.~\eqref{eq:particle density - 1} and \eqref{eq:n_k - 1}.
\item Check convergence of $n_{2m+1}$ and $n_{2m+2}$: if achieved then
set $m\to m+1$ and return to step \ref{enu:block-by-block alg step 2},
else return to step \ref{enu:block-by-block alg step 3} without incrementing
$m$.
\end{enumerate}
The algorithm outlined above is accurate to fourth order in the timestep.
This self-consistent approach is advantageous as one can execute the outer
$\vec{k}$ for-loop in step 3 in parallel which is the most computationally
intensive step of the algorithm. 
The main computational constraint comes from the
time integrals, which require considerable processing and memory resources.
If $d$ is the number of spatial dimensions, $L$ is the number of sites along
a crystal axis, and $N_{t}$ is the number of timesteps, then the memory
requirements scale like $\binom{d + \lfloor L / 2 \rfloor}{d}N_{t}^{2}$.
The binomial coefficient appears as a result of lattice symmetries and the
periodic boundary conditions.
Previous nonequilibrium 2PI studies which integrated similar equations
of motion did not keep all of the history of the memory kernels for
large times, which was justified by the argument that the two-time
correlator would damp at an exponential rate
\citep{Rey1,Aarts1,Aarts2,Aarts3,Aarts4}.
We do not make this assumption since it does not always hold for the
quench protocols we consider. 

\section{\label{sec:Particle number conservation}Particle number conservation}
\begin{widetext}
In this appendix, we identify the terms in the equations of motion
that break particle number conservation. We start with the Dyson's
equation {[}Eq.~\eqref{eq:Dyson's equation - 1}{]} noting that the
bare propagator $G_{0}$ in this context is the atomic propagator
$\mathcal{G}$

\begin{equation}
G_{\vec{k}}^{a_{1}a_{2},c}\left(\tau_{1},\tau_{2}\right)\equiv\mathcal{G}^{a_{1}a_{2}}\left(\tau_{1},\tau_{2}\right)
+\int_{C}\int_{C}d\tau_{3}d\tau_{4}\,\mathcal{G}^{a_{1}a_{3}}\left(\tau_{1},\tau_{3}\right)\Sigma_{\vec{k}}^{\overline{a_{3}}\overline{a_{4}}}\left(\tau_{3},\tau_{4}\right)G_{\vec{k}}^{a_{4}a_{2},c}\left(\tau_{4},\tau_{2}\right).\label{eq:Dyson's equation - 2}
\end{equation}

\noindent Next, we act on both sides with $\delta\left(\tau_{1}^{\prime},\tau_{1}\right)\left\{ i\partial_{\tau_{1}}-E_{\vec{k}}\right\} $,
where for the moment, $E_{\vec{k}}$ is an unspecified function of
$\vec{k}$.  We then integrate over $\tau_{1}$, and set $\left(\tau_{1}^{\prime},\tau_{2}\right)=\left(\tau,\tau^{+}\right)$
and $\left(a_{1},a_{2}\right)=\left(1,2\right)$ to get

\begin{align}
i\frac{\partial}{\partial\tau_{1}}G_{\vec{k}}^{12,c}\left(\tau_{1}=\tau,\tau_{2}=\tau^{+}\right) 
& \equiv E_{\vec{k}}G_{\vec{k}}^{12,c}\left(\tau_{1}=\tau,\tau_{2}=\tau^{+}\right)+\left\{ 
i\frac{\partial}{\partial\tau_{1}}-E_{\vec{k}}\right\} \mathcal{G}^{12}\left(\tau_{1}=\tau,\tau_{2}=\tau^{+}\right)\nonumber \\
 & \quad+\int_{C}\int_{C}d\tau_{3}d\tau_{4}\left\{ i\frac{\partial}{\partial\tau_{1}}-E_{\vec{k}}\right\} \mathcal{G}^{12}\left(\tau_{1}=\tau,\tau_{3}\right)\Sigma_{\vec{k}}^{1\overline{a}}\left(\tau_{3},\tau_{4}\right)G_{\vec{k}}^{a2,c}\left(\tau_{4},\tau_{2}=\tau^{+}\right).\label{eq:Dyson's equation - 3}
\end{align}

\noindent The general form of the contour-time derivative of $G_{\vec{k}}^{12,c}$
is

\begin{align}
\frac{\partial}{\partial\tau_{1}}G_{\vec{k}}^{12,c}\left(\tau_{1},\tau_{2}\right) & =-i\frac{\partial}{\partial\tau_{1}}\left\{ \Theta\left(\tau_{1},\tau_{2}\right)\left\langle \hat{a}_{\vec{k}}\left(\tau_{1}\right)\hat{a}_{\vec{k}}^{\dagger}\left(\tau_{2}\right)\right\rangle _{\hat{\rho}_{i}}^{c}+\Theta\left(\tau_{2},\tau_{1}\right)\left\langle \hat{a}_{\vec{k}}^{\dagger}\left(\tau_{2}\right)\hat{a}_{\vec{k}}\left(\tau_{1}\right)\right\rangle _{\hat{\rho}_{i}}^{c}\right\} \nonumber \\
 & =-i\delta\left(\tau_{1},\tau_{2}\right)-i\Theta\left(\tau_{1},\tau_{2}\right)\frac{\partial}{\partial\tau_{1}}\left\langle \hat{a}_{\vec{k}}\left(\tau_{1}\right)\hat{a}_{\vec{k}}^{\dagger}\left(\tau_{2}\right)\right\rangle _{\hat{\rho}_{i}}^{c}\nonumber \\
 & \quad-i\Theta\left(\tau_{2},\tau_{1}\right)\frac{\partial}{\partial\tau_{1}}\left\langle \hat{a}_{\vec{k}}^{\dagger}\left(\tau_{2}\right)\hat{a}_{\vec{k}}\left(\tau_{1}\right)\right\rangle _{\hat{\rho}_{i}}^{c},\label{eq:derivative of G - 1}
\end{align}

\noindent which also applies to $\mathcal{G}^{12}$. 

The Dyson's equation can also be rewritten as follows

\begin{equation}
G_{\vec{k}}^{a_{1}a_{2},c}\left(\tau_{1},\tau_{2}\right)\equiv\mathcal{G}^{a_{1}a_{2}}\left(\tau_{1},\tau_{2}\right)+\int_{C}\int_{C}d\tau_{3}d\tau_{4}G_{\vec{k}}^{a_{1}a_{3},c}\left(\tau_{1},\tau_{3}\right)\Sigma_{\vec{k}}^{\overline{a_{3}}\overline{a_{4}}}\left(\tau_{3},\tau_{4}\right)\mathcal{G}^{a_{4}a_{2}}\left(\tau_{4},\tau_{2}\right).\label{eq:Dyson's equation - 4}
\end{equation}

\noindent We again act on both sides with $\delta\left(\tau_{2}^{\prime},\tau_{2}\right)\left\{ i\partial_{\tau_{2}}+E_{\vec{k}}\right\} $,
integrate over $\tau_{2}$, and set $\left(\tau_{1},\tau_{2}\right)=\left(\tau,\tau^{+}\right)$,
$\left(a_{1},a_{2}\right)=\left(1,2\right)$ to get

\begin{align}
i\frac{\partial}{\partial\tau_{2}}G_{\vec{k}}^{12,c}\left(\tau_{1}=\tau,\tau_{2}=\tau^{+}\right) & \equiv-E_{\vec{k}}G_{\vec{k}}^{12,c}\left(\tau_{1}=\tau,\tau_{2}=\tau^{+}\right)+\left\{ i\partial_{\tau_{2}}+E_{\vec{k}}\right\} \mathcal{G}^{12}\left(\tau_{1}=\tau,\tau_{2}=\tau^{+}\right)\nonumber \\
 & \quad+\int_{C}\int_{C}d\tau_{3}d\tau_{4}G_{\vec{k}}^{1a,c}\left(\tau_{1},\tau_{3}\right)\Sigma_{\vec{k}}^{\overline{a}2}\left(\tau_{3},\tau_{4}\right)\left\{ i\partial_{\tau_{2}}+E_{\vec{k}}\right\} \mathcal{G}^{12,c}\left(\tau_{4},\tau_{2}=\tau^{+}\right).\label{eq:Dyson's equation - 5}
\end{align}

\noindent Similarly to Eq.~\eqref{eq:derivative of G - 1}, we obtain 
\begin{align}
\frac{\partial}{\partial\tau_{2}}G_{\vec{k}}^{12,c}\left(\tau_{1},\tau_{2}\right) & =i\delta\left(\tau_{1},\tau_{2}\right)-i\Theta\left(\tau_{1},\tau_{2}\right)\frac{\partial}{\partial\tau_{2}}\left\langle \hat{a}_{\vec{k}}\left(\tau_{1}\right)\hat{a}_{\vec{k}}^{\dagger}\left(\tau_{2}\right)\right\rangle _{\hat{\rho}_{i}}^{c}\nonumber \\
 & \quad-i\Theta\left(\tau_{2},\tau_{1}\right)\frac{\partial}{\partial\tau_{2}}\left\langle \hat{a}_{\vec{k}}^{\dagger}\left(\tau_{2}\right)\hat{a}_{\vec{k}}\left(\tau_{1}\right)\right\rangle _{\hat{\rho}_{i}}^{c}.\label{eq:derivative of G - 2}
\end{align}

\noindent It then follows from Eqs.~\eqref{eq:derivative of G - 1}
and \eqref{eq:derivative of G - 2} that

\begin{equation}
\frac{\partial}{\partial\tau_{1}}G_{\vec{k}}^{12,c}\left(\tau_{1}=\tau,\tau_{2}=\tau^{+}\right)+\frac{\partial}{\partial\tau_{2}}G_{\vec{k}}^{12,c}\left(\tau_{1}=\tau,\tau_{2}=\tau^{+}\right)=-i\frac{d}{d\tau_{1}}n_{\vec{k}}\left(\tau_{1}=\tau\right).\label{eq:derivative of n_k - 1}
\end{equation}

\noindent Note that in the special case where $G_{\vec{k}}^{12,c}=\mathcal{G}^{12}$,
one can show explicitly from the analytical expressions for $\mathcal{G}^{12}$
{[}see Appendix C of Ref.~\citep{Fitzpatrick}{]} that the right-hand-side
of Eq.~\eqref{eq:derivative of n_k - 1} vanishes.

Next, by adding Eqs.~\eqref{eq:Dyson's equation - 3} and \eqref{eq:Dyson's equation - 5}
together, summing over all $\vec{k}$, and using Eqs.~\eqref{eq:derivative of G - 1},
\eqref{eq:derivative of G - 2}, and \eqref{eq:derivative of n_k - 1},
we get

\begin{align}
\frac{d}{d\tau_{1}}\left\{ N\left(\tau_{1}=\tau\right)\right\}  & =\sum_{\vec{k}}\int_{C}\int_{C}d\tau_{3}d\tau_{4}\left\{ i\frac{\partial}{\partial\tau_{1}}-E_{\vec{k}}\right\} \mathcal{G}^{12}\left(\tau_{1}=\tau,\tau_{3}\right)\Sigma_{\vec{k}}^{1\overline{a}}\left(\tau_{3},\tau_{4}\right)G_{\vec{k}}^{a2,c}\left(\tau_{4},\tau_{2}=\tau^{+}\right)\nonumber \\
 & \quad+\sum_{\vec{k}}\int_{C}\int_{C}d\tau_{3}d\tau_{4}G_{\vec{k}}^{1a,c}\left(\tau_{1},\tau_{3}\right)\Sigma_{\vec{k}}^{\overline{a}2}\left(\tau_{3},\tau_{4}\right)\left\{ i\partial_{\tau_{2}}+E_{\vec{k}}\right\} \mathcal{G}^{12}\left(\tau_{4},\tau_{2}=\tau^{+}\right).\label{eq:conservation equation - 1}
\end{align}

\noindent Now, if we set $E_{\vec{k}}=\epsilon_{\vec{k}}-\mu$
(i.e we set $E_{\vec{k}}$ to the single-particle excitation energy
of a free particle), and replace $\mathcal{G}^{12}$ by the free propagator for the BHM obtained when $U = 0$, then 

\begin{align}
\left\{ i\frac{\partial}{\partial\tau_{1}}-E_{\vec{k}}\right\} \mathcal{G}^{12}\left(\tau_{1}=\tau,\tau_{3}\right) & \to\delta\left(\tau,\tau_{3}\right),\label{eq:free propagator relation - 1}\\
\left\{ i\frac{\partial}{\partial\tau_{2}}+E_{\vec{k}}\right\} \mathcal{G}^{12,c}\left(\tau_{4},\tau_{2}=\tau^{+}\right) & \to-\delta\left(\tau_{4},\tau^{\prime}\right),\label{eq:free propagator relation - 2}
\end{align}

\noindent and Eq.~\eqref{eq:conservation equation - 1} would become

\begin{equation}
\frac{d}{d\tau_{1}}N\left(\tau_{1}=\tau\right)=\sum_{\vec{k}}\int_{C}d\tau_{3}\left\{ \Sigma_{\vec{k}}^{1\overline{a}}\left(\tau,\tau_{3}\right)G_{\vec{k}}^{a2,c}\left(\tau_{3},\tau^{+}\right)-G_{\vec{k}}^{1a}\left(\tau,\tau_{3}\right)\Sigma_{\vec{k}}^{\overline{a}2}\left(\tau_{3},\tau^{+}\right)\right\} .\label{eq:conservation equation - 2}
\end{equation}

\noindent Baym showed that the term on the right-hand-side of Eq.~\eqref{eq:conservation equation - 2}
vanishes as long as the self-energy $\Sigma$ is of the form $\delta\Phi/\delta G$,
with $\Phi$ a functional of $G$ \citep{Baym,Stefanucci}. As we
mentioned in Sec.~\ref{sec:Equations of motion}, we obtained our
self-energy by taking a functional derivative of the 2PI effective
action, which is indeed a functional of $G$, hence the right-hand-side
of Eq.~\eqref{eq:conservation equation - 2} vanishes and the particle
number is conserved. It is worth stressing that in this scenario, the
self-energy need not be calculated to all orders so that particle number
is conserved. As long as the approximation of the self-energy is of the
form $\delta\Phi/\delta G$, even after taking some low-energy approximation
as we do in our effective theory, conservation will still be guaranteed.

In our case, $\mathcal{G}^{12}$ is not the free propagator for the BHM obtained when $U = 0$, but instead 
is the atomic propagator obtained in the limit when $J=0$. Hence 
there exists no function $E_{\vec{k}}$ in which Eqs.~\eqref{eq:free propagator relation - 1}
and \eqref{eq:free propagator relation - 2} could be possibly satisfied.
The reason for this is due to the asymmetry between the single-particle
and hole excitation energies. For the free propagator, $E^{\left(+\right)}=-E^{\left(-\right)}$,
where $E^{\left(+\right)}$ and $E^{\left(-\right)}$ are the single-particle
and hole excitation energies respectively, whereas for the atomic
propagator $\mathcal{G}^{12}$, $E^{\left(+\right)}\neq-E^{\left(-\right)}$
for all values of $\mu$. Due to this asymmetry, additional terms
are generated leading to

\begin{align}
\frac{d}{d\tau_{1}}N\left(\tau_{1}=\tau\right) & =i\sum_{\vec{k}}\int_{C}\int_{C}d\tau_{3}d\tau_{4}\left[\partial_{\tau_{1}}\mathcal{G}^{12}\right]\left(\tau_{1}=\tau,\tau_{3}\right)\Sigma_{\vec{k}}^{1\overline{a}}\left(\tau_{3},\tau_{4}\right)G_{\vec{k}}^{a2,c}\left(\tau_{4},\tau_{2}=\tau^{+}\right)\nonumber \\
 & \quad+i\sum_{\vec{k}}\int_{C}\int_{C}d\tau_{3}d\tau_{4}G_{\vec{k}}^{1a,c}\left(\tau_{1},\tau_{3}\right)\Sigma_{\vec{k}}^{\overline{a}2}\left(\tau_{3},\tau_{4}\right)\left[\partial_{\tau_{2}}\mathcal{G}^{12}\right]\left(\tau_{4},\tau_{2}=\tau^{+}\right),\label{eq:conservation equation - 3}
\end{align}

\noindent where we introduce the following shorthand notation:

\begin{align}
\left[\partial_{\tau_{1}}\right]G_{\vec{k}}^{12,c}\left(\tau_{1},\tau_{2}\right) & =-i\Theta\left(\tau_{1},\tau_{2}\right)\frac{\partial}{\partial\tau_{1}}\left\langle \hat{a}_{\vec{k}}\left(\tau_{1}\right)\hat{a}_{\vec{k}}^{\dagger}\left(\tau_{2}\right)\right\rangle _{\hat{\rho}_{i}}^{c}-i\Theta\left(\tau_{2},\tau_{1}\right)\frac{\partial}{\partial\tau_{1}}\left\langle \hat{a}_{\vec{k}}^{\dagger}\left(\tau_{2}\right)\hat{a}_{\vec{k}}\left(\tau_{1}\right)\right\rangle _{\hat{\rho}_{i}}^{c},\label{eq:shorthand notation - 1}\\
\left[\partial_{\tau_{2}}\right]G_{\vec{k}}^{12,c}\left(\tau_{1},\tau_{2}\right) & =-i\Theta\left(\tau_{1},\tau_{2}\right)\frac{\partial}{\partial\tau_{2}}\left\langle \hat{a}_{\vec{k}}\left(\tau_{1}\right)\hat{a}_{\vec{k}}^{\dagger}\left(\tau_{2}\right)\right\rangle _{\hat{\rho}_{i}}^{c}-i\Theta\left(\tau_{2},\tau_{1}\right)\frac{\partial}{\partial\tau_{2}}\left\langle \hat{a}_{\vec{k}}^{\dagger}\left(\tau_{2}\right)\hat{a}_{\vec{k}}\left(\tau_{1}\right)\right\rangle _{\hat{\rho}_{i}}^{c},\label{eq:shorthand notation - 2}
\end{align}

\noindent where we now set $E_{\vec{k}}\to0$ as it serves no purpose
for us anymore. The terms on the right-hand-side of
\eqref{eq:conservation equation - 3} are in general not zero.  If we kept all terms 
in the effective theory and did not make the low energy approximation then the the right-hand-side
of \eqref{eq:conservation equation - 3} should equal zero.  However, because the bare propagator 
we use is the atomic propagator Baym's arguments do not hold in the low energy theory and there is 
not conservation of particle number. \end{widetext}

\bibliographystyle{apsrev4-1}
\bibliography{xampl,bibfile}

\end{document}